\documentclass{article}

\usepackage{PRIMEarxiv}
\usepackage[utf8]{inputenc} 
\usepackage[T1]{fontenc}    
\usepackage{hyperref}         
\usepackage{url}                   
\usepackage{booktabs}        
\usepackage{amsfonts} 
\usepackage{mathrsfs}
\usepackage{nicefrac}         
\usepackage{microtype}      
\usepackage{lipsum}
\usepackage{graphicx}
\usepackage{adjustbox}
\graphicspath{ {./images/} }
\usepackage{todonotes}
\usepackage[acronym]{glossaries}
\usepackage{natbib}
\usepackage{caption}
\usepackage{subcaption}
\usepackage{amsmath}
\usepackage{amsfonts}
\usepackage{amssymb}
\usepackage{braket}
\usepackage{float}
\usepackage{alphabeta}
\usepackage{graphicx}
\usepackage{enumerate}
\usepackage{xcolor}
\usepackage{tabularx}

\usepackage{prodint}
\usepackage[titletoc]{appendix}
\usepackage{hyperref}
\usepackage{emptypage}
\usepackage{lscape}
\usepackage{longtable}
\usepackage{color, colortbl}

\usepackage{soul} 
\usepackage{verbatim} 

\title{ Data-Driven Strategies for Detecting and Sampling Misrepresented Subgroups}
\date{}

\author{
 G. Lancia \\
  Department of Mathematics\\
  Università di Genova\\
  Via Dodecaneso, 35. 16146, Genova.\\
  \texttt{giacomo.lancia@gmail.com} \\
   \And
   F. Mecatti \\
  Department of Sociology and Social Research\\
  Università degli studi di Milano-Bicocca\\
  Via Bicocca degli Arcimboldi, 8. 20126 Milano\\
\texttt{fulvia.mecatti@unimib.it}\\
\And
    E. Riccomagno \\
  Department of Mathematics\\
  Università di Genova\\
  Via Dodecaneso, 35. 16146, Genova.\\
\texttt{riccomag@dima.unige.it} \\
}

\newacronym{mse}{MSE}{Mean Square Error}
\newacronym{rb}{RB}{Relative Bias}
\newacronym{ann}{ANN}{Artificial Neural Network}
\newacronym{vae}{VAE}{Variational Auto-Encoder}
\newacronym{pca}{PCA}{Principal Component Analysis}
\newacronym{kpca}{KPCA}{Kernal Principal Component Analysis}
\newacronym{mcc}{MCC}{Matthews Correlation Coefficient}
\newacronym{ae}{AE}{Auto-Encoder}
\newacronym{kld}{KLD}{Kullback-Leibler Divergence}
\newacronym{jsd}{JSD}{Jensen-Shannon Divergence}
\newacronym{mf}{MF}{Multi-Frame}
\newacronym{sts}{STS}{Stratified Sampling}
\newacronym{srs}{SRS}{Simple Random Sampling}
\newacronym{eusilc}{EU-SILC}{European Union Statistics on Income and Living Conditions}
\newacronym{ehi}{EHI}{Equivalise Household Income}
\newacronym{ml}{ML}{Machine Learning}
\newacronym{sm}{SM}{Simple Multiplicity}
\newacronym{pml}{PML}{Pseudo-Maximum Likelihood}

\definecolor{Kol2023}{RGB}{174,193,197}

\begin{document}
\maketitle

\begin{abstract}
Economic policy research often investigates population well-being, with particular attention to the links between unequal living conditions, low education, and social exclusion. Sample surveys such as EU-SILC are widely used for this purpose and inform public policy, yet their sampling designs may fail to adequately represent rare, hard-to-sample, or under-covered subgroups. This limitation can hinder socio-demographic analyses and evidence-based policy design. We propose a generalisable approach based on univariate and multivariate unsupervised learning techniques to detect outliers in survey data that may signal under-represented subgroups. Identified groups can then be characterised to inform targeted resampling strategies that improve survey inclusiveness. An empirical application using the 2019 EU-SILC data for the Italian region of Liguria shows that citizenship, material deprivation, large household size, and economic vulnerability are key indicators of under-representation. 
\end{abstract}

\section{Introduction}\label{sec: intro}

As our world becomes increasingly interconnected, the role of poverty and inequalities in wealthy societies becomes ever more important.
To capture and monitor the complex, intricate evolution of social dynamics, European countries participate in the European Union Statistics on Income and Living Conditions programme (\gls{eusilc})~\cite[]{EUSILC:2019}, 
which is the reference source for comparative statistics on income distribution and social inclusion in the European Union and is used for policy monitoring within the Open method of coordination~\cite{glossary-eusilc}.
This paper focuses on users of ~EU-SILC data who are committed to representation equity and aim to produce equitable statistics. 
For~EU-SILC being a survey intended for general users and purposes, it may pose challenges in both respects. 
Representation equity is a key objective of official statistics and a necessary condition for data-driven and unbiased policy-making. It involves fostering the visibility of misrepresented groups in the data, which may hide under-served communities~\cite[]{NASEM2023}. 
This paper is concerned with  two  main objectives:
\begin{enumerate}
\item 
To apply advanced data-driven methods to EU-SILC cross-sectional data in order to detect hidden, rare, or otherwise under-represented population subgroups ---such as vulnerable or marginalised communities--- that may require targeted policy attention. In practice, this concerns survey sampling challenges related to rare or hard-to-reach subpopulations, response behaviour and declining participation rates (since vulnerable groups often have a lower propensity to respond), and the possible existence of unknown or unplanned subgroups that are relevant but under-represented in the sample data.

\item To discuss state-of-the-art sampling strategies for the effective integration of~EU-SILC data, specifically targeting misrepresented subgroups and populations segments not sufficiently captured in the EU-SILC sample.
\end{enumerate}

Both objectives relate to critical concerns,  specifically, hard-to-sample populations and undercoverage of the sampling frame, that are well known in survey sampling methodology and practice and have produced a consistent and well-established body of literature (see, for instance, popular reviews such as~\cite{kalton2009methods,tourangeau2014hard}. 
Although such literature can help address objective 2 above, a literature gap emerges with respect to objective 1. The purpose of detecting misrepresented population subgroups may, in fact, extend beyond known and pre-defined subgroups, for instance, strata of the general population not included as eligible for the survey, to also encompass unknown and hard-to-predict subgroups that could still be of interest. 
Recent literature addressing this issue focuses on health studies (see, for instance,~\cite{tornava2025hard}), and more methodological research is needed in the context of social research. 
This paper contributes to filling this literature gap, specifically targeting both known and {\em a priori} unknown misrepresented subgroups in~EU-SILC datasets. However, we believe that both the objectives and the methodological strategies proposed in this paper have wider applicability to different contexts and datasets, with the potential to contribute more broadly to addressing this under-researched issue.

This paper is motivated by a real case study proposed by the local government of the Italian Liguria region (see sub-section~\ref{sec: liguria_project}). 
As a consequence, the objectives above also entail an improved granularity in order to reflect geographical heterogeneity, which is especially relevant in a country like Italy and for variables as measured in ~EU-SILC. 
A high-quality large-scale probability survey of households, as is the case for~EU-SILC, ensures accurate estimates at the global EU level and for each nation as a whole. 
However, the European and national large sample sizes can be inadequate to meet the data demands at the sub-national level and for key population subgroups, which can result in data inequities with subsequent lack of information for unbiased policy making. 

The remainder of this article is organised as follows: 
Section~\ref{sec: data} provides a concise description of the dataset under investigation, namely the Ligurian subsample of the 2019~EU-SILC microdata. 
This section also addresses the issue of misrepresented subgroups in the~EU-SILC data, particularly in Liguria, together with the reasons why this lack of representation was expected, as well as the sampling design context of the~EU-SILC survey in the region. 
Section~\ref{sec: methodology} describes the methodology employed, while Section~\ref{sec:Results} presents the results obtained. 
To mitigate the problem of missing information of subgroups, in Section~\ref{sec: Intergative_Sampling} we present a simulation study to evaluate integrative sampling schemes to append to the existing survey. 
Finally, comments and a broader discussion of the entire process are provided in Section~\ref{sec: discussion}.
The supplementary material includes Appendix~\ref{apx: additional_info_data}, which provides detailed information on the EU-SILC variables used in the analysis, and Appendix~\ref{apx: Entropy_Score}, which presents additional technical details intended for readers without a background in supervised learning, while for a background on the multi-frame approach we refer to~\cite{mecatti2007single}. 
Finally, Appendix~\ref{apx:simulation_allocation} details the simulation settings used in Section~\ref{sec: Intergative_Sampling} and provides information on the availability of the code for the entire article.

\section{Background, Context and Data}\label{sec: data}

Despite the proliferation of new technologies, poverty and inequality persist in European societies, and the anticipated positive effects of technological advancement on these socioeconomic issues have not materialised as expected~\cite[]{atkinson1998world,longford2014statistical}.
In opposition to this, the role of education in contrasting poverty and inequality has been largely proven~\cite[]{hofmarcher2021effect, raffo2007education}. 
Among countries in the~EU-SILC programme, low-level education has been correlated with a gap in income that  affects children's educational outcomes~\cite[]{ferguson2007impact}. 
In addition, integration plays an important role in comprehending social inequalities, such as the impact of unskilled immigrants on natives~\cite[]{gang1994labor} and the difficulty these immigrants experience in owning a home or earning higher incomes once migrated with few resources~\cite[]{lewin2023poverty}.

When considering the use of the EU-SILC microdata for monitoring the current living conditions in Europe, it is crucial to critically assess the quality of these data. 
Recent studies have shown that the EU-SILC microdata may vary significantly in its depiction of socio-economic conditions across European countries. 
This discrepancy is particularly noticeable in income-related variables~\cite[]{trindade2020comparability}.
Moreover, country-specific issues regarding the quality of the samples collected through the~EU-SILC programme have been identified as early as 2008, particularly in the German section~\cite[]{hauser2008problems}.
A 2020 analysis by~\cite{eurostat2020comparability} revealed substantial compliance issues with Eurostat’s guidelines in several member states, exacerbating the challenges in achieving cross-national comparability. 
These findings underscore the need for increased efforts to ensure the accuracy and consistency of the~EU-SILC dataset.
Major challenges might also arise when such an open-access source of data is utilised to support local policies, which highlights a need for more robust approaches to improve representation equity, as discussed above. 
It is worth noting that the~EU-SILC programme and its sampling scheme remained unchanged up to 2023. 
A recent upgrade to the questionnaire was implemented in 2024 to provide a more reliable and consistent source of data~\cite[]{wirth2022european}. 

\subsection{The Liguria Project}\label{sec: liguria_project}

This study was initiated in response to an internal need of the Liguria Region to evaluate the impact of newly implemented policies, primarily aimed at promoting a \emph{Pact for Residential Stability} and \emph{Family Support} measures. 
These initiatives were designed to reduce the fiscal burden on households and stimulate investment and consumption; see art.~5 of~\cite[]{liguria2017legge}.
The policies were proposed in 2017, approved in 2018, and introduced in 2019. 
Specifically, for that year, tax exemptions were granted to large families and citizens with a low or medium-low taxable income, provided they had at least one dependent child born in 2018~\cite[]{liguria2017legge}.
In assessing the effectiveness of these measures, the Region faced a lack of specific internal data and found the EU-SILC 2019 Liguria dataset insufficient for this purpose, as those households were not adequately represented in the sample. Consequently, it prompted this study to examine in greater detail the case of under-represented or misrepresented populations within the EU-SILC cross-sectional sample specific for Liguria.
In order to develop an efficient rich plan of assistance services and extend them to those groups of the resident population that are more exposed to inequality and poverty factors, referred to as vulnerable, frail and marginalised residents, the administrative bureau of the Liguria region has recently proposed to fill the gap in the available data, particular in the Ligurian sub-sample of the~EU-SILC programme, on their specific challenges.
Our work has the ultimate scope to address this gap.

\subsection{2019 EU-SILC: Sample Design and Data}\label{sec: Sample Design and Data}

The main purpose of the~EU-SILC programme is to collect timely and comparable cross-sectional and longitudinal multidimensional micro-data on income, poverty, social exclusion and living conditions, produced annually and covering all 27 EU Countries plus Iceland and Norway~\cite[]{eusilc2019methodology, EUSILC:2019}.
The general guidelines for the 2019~EU-SILC survey rigorously define the reference population and eligibility criteria as 
\begin{quote}
    All private households and their current residing members at the time of data collection, excluding collective households and institutions. 
    All private households and all persons aged 16 and over within the household are eligible for the survey operation 
\end{quote} 

Children under the age of 16 are covered indirectly through household-level information reported by adult members.
At the same time, the guidelines grant the National Statistical Offices of all participating Member States in the~EU-SILC programme, who are responsible for data collection, considerable flexibility in their choice of sampling design.
This flexibility is exercised with a focus on delivering timely and internationally comparable cross-sectional data, while implementing appropriate procedures, in line with each office’s own best practices, to maximise response rates.
 
A nationally representative probability sample of the target population is required,  both at the household and at individual person level. 
The sampling frame and sample selection ensure a known and positive (first-order) inclusion probability. 
The national minimum effective sample size is stated based on practical considerations and precision requirements for the released estimates, e.g. for Italy 2019, a cross-sectional sample of  $7250$ households and of $15500$ persons aged $16$ and over to be interviewed~\cite[]{eusilc2019methodology}.

ISTAT (the Italian official statistics agency) in its~EU-SILC 2019 methodological report~\cite{ISTAT2019}, guarantees a sample size of (circa) $22000$ households and $50000$ individuals across $650$  municipalities (Comuni), leading to a national sample fraction of  8.4 per 10000 individuals (0.084\%). 
In addition to the EU general guidelines requests, ISTAT also assures representative  (cross-sectional) probabilistic samples and precise estimates at the regional level. 
No minimum effective sample size is prescribed for the Ligurian cross-sectional sample, as sampling is primarily conducted at the municipality level rather than directly at the regional level. Specifically, the 2019 Ligurian cross-sectional sample comprises $1925$ individuals from 990 households across 21 municipalities, corresponding to an individual sample fraction of 0.3 per 10,000 (0.003\%) of the Italian population and 13 per 10,000 (0.13\%) of the Ligurian population.
The EU-SILC Italian sample design has two stages with stratification of the primary sample units (PSU) represented by Comuni and secondary sample units (SSU) represented by households. The first stage sample is selected with probability proportional to PSU demographic size and includes: the largest PSUs that inform a unique stratum, four PSUs selected within strata informed by at least four Comuni; and, otherwise,  two selected PSUs. The second stage sample is a simple random without replacement (SRS).
According to the ISTAT methodological notes~\cite{ISTAT2019} collected data are processed and edited; 
missing data imputed,  also on the basis of external data from administrative official registries;
original design weights are adjusted for item non-response and according to two steps of calibration to available known total controls.
Access to aggregated data is open to the general public. Microdata can be accessed on demand for scientific purposes and projects \url{(https://ec.europa.eu/eurostat/web/microdata/european-union-statistics-on-income-and-living-conditions)}.

The~EU-SILC database offers a rich array of information of various kinds.
Our case study focuses on a subset of variables selected from the questionnaire concerning the living conditions and social exclusion of Liguria's population. Specifically, to prioritise the identification of macroscopic trends, only variables uniformly applicable across the population were considered. Variables that could introduce structurally missing data, due to being limited to specific respondent categories, such as pensioners, renters, households with minors, or recipients of employment-related benefits, were excluded. The resulting dataset is complete, requiring no imputation or adjustment for missing data. This led to the selection of 19 discrete variables (ordinal, categorical, and dichotomous), while quantitative and continuous variables self-reported and required discretisation were omitted.
A full description of the selected variables is provided in Appendix~\ref{apx: additional_info_data}.

To conclude this subsection, we anticipate the core contribution of this work, presented in  Sections~\ref{sec: methodology} and~\ref{sec:Results}. It addresses objective 1 of Section~\ref{sec: intro}, through the identification of anomalies and outliers within the empirical distribution functions of the selected study variables.
This intuition was prompted by a comparison of the representativeness of~EU-SILC data with respect to a complete administrative source of socio-demographic data concerning tax records.

\subsection{Liguria Complete Tax Administrative Data} \label{sec: data_tax}  

\subsubsection{2019 Tax Declaration Dataset}\label{sec: data_tax_decalration}
From the Italian national tax records warehouse, we extracted the portion of specific administrative information concerning the Ligurian region in 2019, hereafter referred to as the \emph{Tax Declaration Dataset}. 
This dataset is non-public, subject to data protection regulations, and includes anonymised tax records for the Ligurian population.
It offers a comprehensive overview of the income and economic status of most households in the region. 
It contains a total of 1.216.747 individual tax declarations, corresponding to 731.537 distinct households. 
When compared to the official population census data provided by ISTAT, this coverage accounts for approximately $96.5\%$ of all Ligurian households, making it a highly representative and detailed data source.
Because of its quasi-complete coverage, this dataset provides reliable demographic and economic indicators, such as sex, age, employment status, annual income, number of dependent children, and number of dependent disabled children. 
We therefore employed it as a benchmark for assessing the representativeness of selected variables within the Ligurian~EU-SILC cross-sectional sample.

\subsubsection{Comparative Analysis of EU-SILC and Tax Declaration Dataset}\label{sec: comparative_analysis_EUSILC_TAX_DECALARATION}
\begin{figure}[h]
  \centering
  \begin{subfigure}{0.48\textwidth}
    \includegraphics[width=\linewidth]{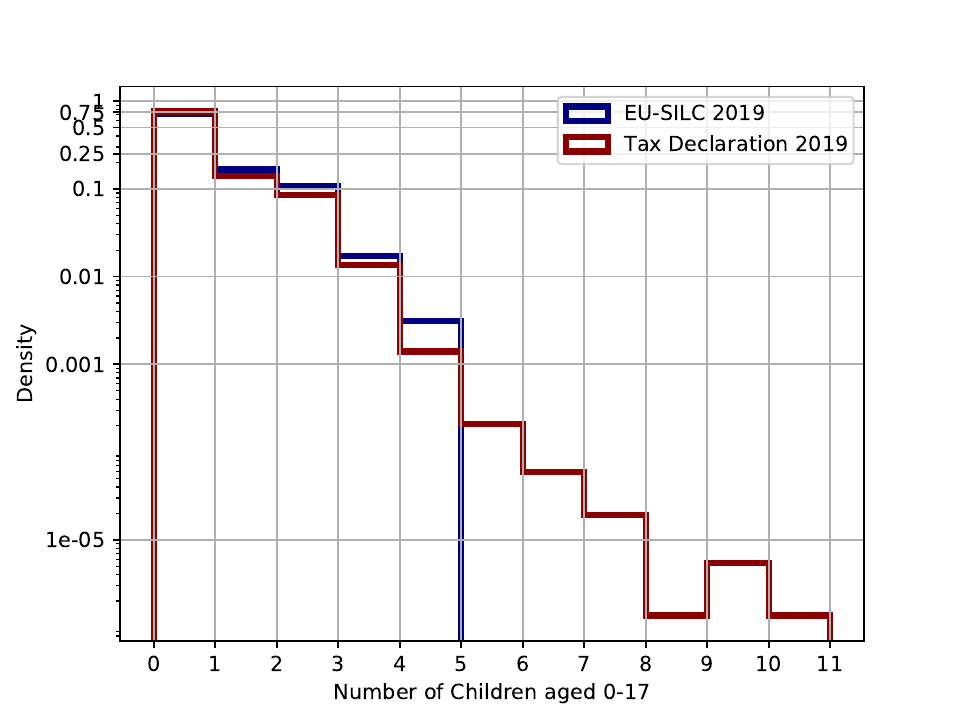}
    \caption{Children aged 0-17}
  \end{subfigure}
  \hfill
  \begin{subfigure}{0.48\textwidth}
    \includegraphics[width=\linewidth]{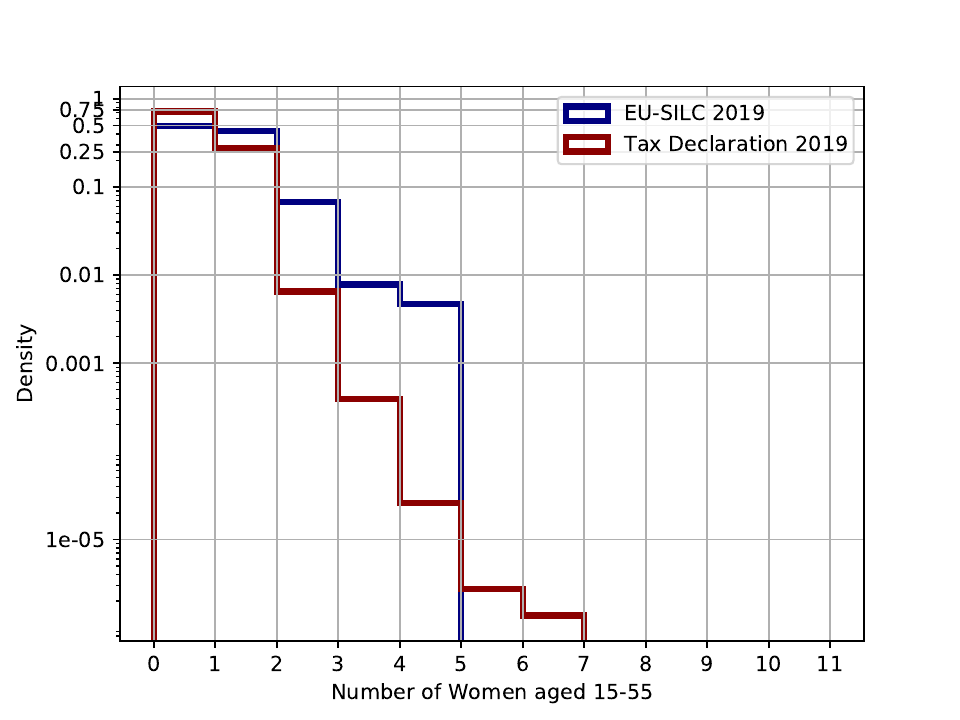}
    \caption{Women aged 15-55}
  \end{subfigure}
  \caption{Comparison of household-level distributions for children aged 0-17 and women aged 15-55, in the Ligurian subsample of the 2019~EU-SILC survey and the administrative Tax Declaration Dataset for Liguria (2019).
  The graphs illustrate discrepancies between the two data sources and potential coverage issues in~EU-SILC sample data.}
  \label{fig: comparision_EUSILC_TAX_DECLARATION}
\end{figure}

We leverage the Tax Declaration Dataset to compare against the sampled data collected by~EU-SILC, aiming to highlight potential mismatches and identify demographic subgroups that may be under-represented or missed altogether in the survey. 
To this end, we focus on two variables central to our analysis and available in both datasets: the number of women aged 15-55 and the number of children aged 0-17 per household.
In Figure~\ref{fig: comparision_EUSILC_TAX_DECLARATION} we present comparative histograms that visualise differences in distribution and coverage between the tax-based and survey-based sources. 
The~EU-SILC sample design is taken into account by using (adjusted) design weights, see Section~\ref{sec: Incorporating Sample Design Weights}. 

As shown in Figure~\ref{fig: comparision_EUSILC_TAX_DECLARATION}, there are notable discrepancies between the two data sources.
For children aged 0-17, the~EU-SILC data fail to capture households with more than five children,  an extremely hard-to-reach subpopulation with a prevalence of approximately $0.1\%$. Conversely, for households with one to four children, the~EU-SILC appears to slightly over-represent these cases compared to the tax data.
Similarly, for women aged 15-55, the~EU-SILC dataset underrepresents households with more than five women in this age range, a highly rare group accounting for at most $0.001\%$ of the Ligurian population. 
At the same time,~EU-SILC tends to slightly over-represent households containing between two and five women in this age range. Notably, households with no women aged 15-55 are substantially underrepresented in~EU-SILC, comprising only about 50\% of the sample compared to 72\% in the Tax Declaration dataset.
Moreover, some relevant subgroups, which are well captured in administrative datasets, are entirely absent from the~EU-SILC dataset. 
For instance, the Tax Declaration dataset includes information on dependent children with disabilities, a population segment that was not investigated in the 2019~EU-SILC survey.

These findings highlight emerging examples of underrepresented subgroups in the~EU-SILC data. This representation bias may result either from the inherent difficulty in reaching rare population subgroups, e.g., large households with many children or multiple women of working age, or from ~EU-SILC  frame under-coverage issues, e.g., households with no women aged 15-55). 
This evidence supports the idea that the identification of such discrepancies can be effectively approached as an outlier detection problem. 
This perspective may help bring to light groups systematically under-detected by~EU-SILC’s sampling design.

\section{Detecting misrepresented subgroups : the proposed methodology}\label{sec: methodology}

This section describes the methodology employed to identify the misrepresented subpopulation of interest. 
The proposed methodology involves adapting three \emph{outlier detection} techniques to identify segments of the Ligurian population that may constitute misrepresented subgroups.
Operatively, we work with unlabelled data and aim to construct a binary labelling that separates \emph{typical observations} from outliers.
Typical observations are defined as the most recurrent patterns observed across the majority of the EU-SILC data, whereas outlier (in the sequel also called \emph{atypical}) observations are those that occur infrequently.
It is this lack of recurrence that constitutes the core focus of our analysis.
At this stage, we do not aim to explain the nature of the anomalies, but rather to highlight them as indicators of potential information gaps, serving as red flags for areas of possible misrepresentation requiring further investigation.

The methodology we propose necessarily contends with the absence of ground truth, which limits our evaluation to internal validation procedures.
Consequently, comparing the performance of different outlier detection models does not serve to verify the correctness of a given label, but rather to assess the consistency with which each model identifies certain observations as atypical.
In this context, an observation flagged as an outlier should be interpreted as a strong indication, emerging across multiple learning folds and model runs, that the data point exhibits patterns systematically distinct from the majority.

\subsection{Anomaly detection models}\label{sec: Anomaly detection models}
We chose to implement models to include both univariate and multivariate approaches: the former analyses variables individually, while the latter accounts for their interconnections.

As a unidimensional strategy, we propose an \emph{entropy-based score}, which exploits the property of entropy as an aggregate measure of information, typically suggesting how infrequent an observation is~\cite[]{carter2007introduction}.
The rationale for including this strategy in our study lies in its ability to analyse each variable independently. 
Specifically, such a strategy allows us to examine the intrinsic atypicality of individual variables, treating them as unassociated. 
As a result, the identification of misrepresented subpopulations consists of aggregating those variables that both predominantly influence the anomaly detection process and stand out as markedly atypical.

For a multivariate perspective, we investigated both standard statistical learning approaches and~\gls{ml}-based methods.
Specifically, we focused on reconstruction-based models, i.e.,~\gls{kpca} and~\gls{ae}.
They share a common principle: to encode high-dimensional data into a lower-dimensional latent space and subsequently reconstruct the original input through an inverse transformation, either exact or approximate.
The encoding and decoding procedures are typically derived through optimisation processes that seek to span data into a low-dimensional encoding domain while preserving as much of the original variance as possible. 
In the context of anomaly detection, the reconstruction error itself serves as a signal: observations that cannot be adequately reconstructed, i.e., those with high reconstruction error, are flagged as potential outliers, because their patterns deviate from those learned from the bulk of the data.

\subsubsection{Entropy Score}\label{sec: Entropy Score}   

We leverage entropy within a univariate framework, treating dataset variables individually in order to isolate distinct patterns that effectively highlight potentially misrepresented subgroups.
Our approach consists of two main steps: first, we compute an entropy-based score for each variable to identify those most informative and prone to containing outliers; second, we use the concept of information content derived from entropy to pinpoint the specific categories within these variables that best represent the target anomalies.

For a given categorical variable $X$ observed in the sample with $k$ categories $x_i$ and  relative frequency  $p_i$,  we considered the following score based on Shannon Entropy Index~\cite[]{mackay2003information}
\begin{equation}
   \Gamma_1 = 1 + \frac{1}{\log k} \sum_{i=1}^{k} p_i \log p_i,
\end{equation}
The score $\Gamma_1$ ranges between 0 and 1; higher values indicate a greater likelihood of detecting misrepresented categories. After computing $\Gamma_1$ for our focus set of 19 variables (see Appendix~\ref{apx: additional_info_data}), we select the top five most informative ones (see Section~\ref{sec: Outlier_Detection}).
For each of these variables, we then assess the information content of each category using $-\log{p_i}$, which helps pinpoint potentially misrepresented groups within those variables.
Additional technical details can be found in the Appendix~\ref{apx: Entropy_Score}.

\subsubsection{Kernel Principal Component Analysis}\label{sec: KPCA}

A non-linear extension of~\gls{pca} is~\gls{kpca}~\cite[]{mika1998kernel,scholkopf1998kernel}, which generalises the classical approach by implicitly mapping the original data into a high-dimensional feature space, where linear~\gls{pca} is then performed~\cite[]{hastie2009elements}.
The core idea behind~\gls{kpca} for anomaly detection is that it performs a nonlinear transformation of the data into a higher-dimensional feature space via a mapping function $\Phi$. 
This transformation, learned from the data, extends the traditional~\gls{pca} framework by capturing nonlinear structures. 
Following the strategy usually adopted for~\gls{pca}, data are projected into a higher-dimensional space (in this case, the $\Phi$-space). 
Observations that exhibit poor reconstruction quality, i.e., a large \emph{reconstruction error}, are likely to be outliers, as they do not conform well to the learned data patterns.
The \emph{reconstruction error} in the feature space is evaluated as 
\[\operatorname{RE}_{\text{KPCA}}(\Phi(\mathbf{z})) = \| \Phi(\mathbf{z}) - \hat{\Phi}(\mathbf{z}) \|^2\]
where $\mathbf{z}$ is the original data point and $\hat{\Phi}(\mathbf{z})$ denotes the projection of $\Phi(\mathbf{z})$ onto the subspace capturing 95\% of the variance in the transformed feature space.
To determine the transformation 
$\Phi$, one typically employs kernel matrices, which implicitly embed the desired transformation. In our case, we utilise a Hamming metrics-based kernel matrix, ensuring that the constructed kernel is positive semidefinite. Subsequently, as for~\gls{pca}, the eigen decomposition of the kernel matrix enables the determination of the transformations required for decomposing and reconstructing the data.
Further technical details are provided in Appendix~\ref{apx: KPCA}.

\subsubsection{Auto-Encoder}\label{sec: AE}

An Auto-Encoder~(\gls{ae}) is a neural network designed to learn efficient representations of input data in an unsupervised manner~\cite{aggarwal2018neural}. 
It consists of two components: an \emph{encoder} and a \emph{decoder}, which are typically designed as two feed-forward neural networks (usually, multi-layer perceptrons) with the scope of compressing and subsequently decompressing data. 
Unlike other reconstruction-based methods, neural networks have the advantage of being universal approximators of any measurable function~\cite[]{calin2020deep}, which establishes the superiority of~\gls{ae}-based methods in exploring optimal embedding transformations to make unsupervised learning efficient~\cite[]{aggarwal2018neural, bishop2006pattern}.  
Formally, let us consider a data matrix, $X \in \mathbb{R}^{n\times p}$, where 
$n$ denotes the number of samples (observations), 
$p$ denotes the number of features (variables) measured for each sample and the $(i,j)$ element of $X$ is the value of the feature $j$ observed for the $i$-th sample. A generic row of $X$ is indicated by $\mathbf{x}$. 
An encoder is defined as a map $f_\theta: \mathbb{R}^p \to \mathbb{R}^d$ that maps the input $\mathbf{x} \in \mathbb{R}^p$ to a latent representation $\mathbf{z} \in \mathbb{R}^d$; usually $d<p$, and so $\mathbf{z}$ contains a smaller, simplified version of the information in $\mathbf{x}.$
Likewise, a \emph{decoder} is a map $g_\varphi: \mathbb{R}^d \to \mathbb{R}^p$ that reconstructs the input from the latent representation.
The subscripts $\theta$ and $\varphi$ denote the sets of parameters learned in the hidden layers of the encoder and decoder, respectively.
Such a model is trained to minimise the reconstruction error, which represents how the compression-decompression mechanism deviates from the input data, considered here as a benchmark. 
Put simply, the~\gls{ae} learns a sort of optimal transform along with its inverse, to reduce the dimensionality of input data; for this reason, this method represents a wide use adaptive strategy for data reduction in unsupervised learning.
Typically, the reconstruction error is expressed as:
\[
\mathcal{L}_{\text{AE}}(\mathbf{x}; \theta, \varphi) = \frac{1}{n}\sum_{j = 1}^{n}\| \mathbf{x}_j - g_\varphi \left( f_\theta(\mathbf{x}_j) \right) \|^2.
\]
For the outlier detection problem,~\gls{ae}-based strategies have been proposed in several applications 
\cite[]{abhaya2023efficient, an2015variational, du2022graph, torabi2023practical, zhou2017anomaly}, among others. 
The underlying assumption is that the~\gls{ae} learns how to accurately compress and reconstruct \emph{typical} data but will yield a high reconstruction error for anomalous inputs not well represented in the training distribution. 
Thus, the reconstruction error serves as an unsupervised anomaly score.

In this work, we designed an~\gls{ae} following general rules of thumb, as suggested in~\cite{aggarwal2018neural, Goodfellow-et-al-2016}.
To avoid overfitting and unnecessary complexity, we opted for a shallow architecture: both the encoder and decoder consist of two hidden layers each.
The encoder's first layer is a dense (fully connected) layer with a hyperbolic tangent activation function, which reduces the dimensionality of the input data by half. The second layer, identical in structure, further reduces the dimensionality by half. Consequently, the embedding space has a dimensionality equal to one-quarter of the original input.
The decoder mirrors the encoder in structure. 
The first layer is a dense layer that doubles the size of the embedding vector. The second layer is also dense and uses a softplus activation function~\cite[]{glorot2011deep}, which smoothly transforms the outputs and restores the vector to its original dimensionality. 
In other words, the second layer applies a smoothing nonlinear function, i.e., the softplus function, defined as $\log(1 + \exp(\cdot))$, to ensure all outputs remain positive while keeping the overall structure of the data
Finally, since the input data are discrete, we include an additional rounding step when evaluating the reconstruction error.
This post-processing step ensures consistency with the original data format while preserving the benefits of smooth training dynamics.

\subsection{Outlier Detection}\label{sec: Outlier_Detection}

So far, we have introduced the methodologies in terms of the scores they compute for outlier detection. 
In this section, we describe our approach for identifying outliers within the specific context of the methods presented above.  
The key issue we address is that, while both the Entropy-Based Score and the reconstruction error-based methods provide meaningful scores (with higher values indicating a higher likelihood of being an outlier), they do not supply a threshold for classifying observations as outliers. 
To overcome this issue, we adopt an unsupervised strategy.  
For the Entropy-Based Score (see Section~\ref{sec: Entropy Score}), we apply $k$-means clustering with $k=2$ directly to the score distribution. 
This yields a data-driven threshold on $\Gamma_1$, enabling us to identify variables that may contain misrepresented subpopulations.  
For the~\gls{kpca} (see Section~\ref{sec: KPCA}) and~\gls{ae} (see Section~\ref{sec: AE}) methods, we proceed analogously, but at the instance level rather than the variable level. Specifically, we apply $k$-means clustering with $k=2$ to the reconstruction error scores, because items with higher reconstruction errors are more likely to be outliers.
In fact, clustering based on a one-dimensional score, whether $\Gamma_1$ or a reconstruction-based error, separates the data into two groups, with the cluster having the higher centroid (i.e., larger average score) labelled as outliers.

\subsection{Stability Validation}\label{sec: Stability_Validation}

Following~\cite{ben2002pacific, lange2004stability}, and given the unsupervised nature of our setting where no ground truth is available, we applied a stability validation procedure to assess the robustness of the clustering outcome. Stability validation evaluates whether small perturbations in the data lead to consistent clustering solutions, thereby indicating that the detected structure is reliable rather than an artefact.
In our case, we adopted a leave-one-out strategy: at each iteration, one item was excluded and clustering was performed on the remaining items. 
The rationale is that removing a single element should not substantially alter the overall clustering structure if it is stable. To quantify this, we compared the cluster solutions obtained across iterations using the~MCC~\cite{matthews1975comparison}. Although primarily applied in supervised learning,~MCC can also be used to provide a balanced measure of agreement between clustering assignments~\cite[]{chicco2023statistical}. 
It measures the consistency between cluster labels derived from cross-validation and those obtained from the full-data model. 
A high MCC value indicates that the clustering procedure reliably identifies the same outlier characteristics across different partitions of the data, thus supporting the robustness of the approach.
This procedure allowed us to evaluate the robustness of the 2-cluster K-means algorithm in consistently identifying outlier structures across the dataset.

\subsection{Internal Validation}\label{sec: Internal_Validation}

To complement stability validation, we performed an internal validation using a 10-fold cross-validation procedure. 
Internal validation aims to verify whether the clustering outcome remains consistent when the data are partitioned into subsets, thereby assessing the generalisability of the method.
At each iteration, one fold was excluded while the remaining nine folds were used to train the model and perform clustering. 
The excluded fold then served as a validation set for outlier detection. 
Because the folds are non-overlapping, this process generated a collection of clustering assignments for the excluded data. 
We compared these assignments against those obtained by training the model on the full dataset, under the assumption that removing a single fold should not substantially alter the underlying clustering structure. 
A robust method is therefore expected to produce similar outcomes across folds.

\subsection{Variable Inspection}
To extrapolate from the proposed multivariate unsupervised methodology, which variables (and the corresponding population subgroups) contribute most to the anomalous characteristics of the detected outliers, we employed the \emph{Feature Importance} algorithm introduced by~\cite{breiman2001random}.
This method was originally presented to inspect supervised models.
It operates by disrupting the relationship between each input variable and the target variable, assessing the extent to which the 
model’s predictions degrade when the association is broken.
In our application, each variable is shuffled one at a time, while all other variables remain unchanged.
The modified data are then fed into the anomaly detection models, and their scores are passed to the fitted clustering model.
The impact due to the shuffling of a variable is assessed by measuring the \emph{importance score}, namely, the change in performance, quantified as the difference in the \emph{accuracy score} (i.e., the fraction of correct predictions) for the identification of anomalous items.
This difference is used as an estimate of the variable’s importance.
The greater the importance assigned to a variable, the more influence it exerts on the outlier detection process.
To ensure stability, each variable was permuted 30 times.
The mean importance score and the corresponding 95\% confidence interval for each feature were estimated via jackknife resampling.
This approach provides a robust quantification of each feature’s influence on the detection of anomalies.

\subsection{Exploring Under-represented Subgroups}
To identify potential under-represented subpopulations among the items previously labelled as outliers, we perform an additional clustering step restricted to this subset. 
Given that the structure of the data may not reside in a convex space, we adopt spectral clustering, a technique specifically suited for identifying complex, non-convex cluster geometries~\cite[]{hastie2009elements}.
It is important to note that this secondary clustering procedure is applied exclusively to the outliers identified by the unsupervised anomaly scores derived from the~\gls{kpca} and~\gls{ae}. 
The spectral clustering is executed solely on the set of outlier items, with the aim of discovering internal structure or subgroupings within this anomalous region.
To interpret the resulting clusters, we focus on visualising their most representative members, specifically, the \emph{medoids} of each cluster. 
The medoid is defined as the observation within a cluster that minimises the average distance to all other points in the same cluster, thereby offering an interpretable exemplar of the cluster’s characteristics~\cite[]{struyf1997clustering}. 
The optimal number of clusters is determined by selecting the configuration that maximises the \emph{Silhouette Score}~\cite[]{rousseeuw1987silhouettes} over a predefined range of possible cluster counts. 
For consistency with our earlier methodology, we construct the affinity matrix for spectral clustering using the same Hamming distance-based kernel employed in the~\gls{kpca}. 
The Hamming distance is also used in the computation of silhouette scores. 

\subsection{Incorporating Sample Design Weights}\label{sec: Incorporating Sample Design Weights}

When employing the~EU-SILC data to train the anomaly detection models, it is essential to account for the sample design by means of the proper weighting system.
%
As discussed in Section~\ref{sec: Sample Design and Data}, the~EU-SILC survey adopts a two-stage sampling design in which Comuni serve as PSUs and households as SSUs. Together with the microdata, Eurostat provides design weights calibrated (in the following simply named weights) to available population totals to account for non-response and to enhance estimation efficiency. Weights can be directly applied to training sets and incorporated into the reconstruction or loss function.
We then focus on the Horvitz-Thompson estimate of the mean of a generic 
variable $X$ computed over a sample of size $n$ by using those weights.
In this way, the loss function remains consistent with the underlying survey design, ensuring that the anomaly detection model learns patterns that reflect population-level characteristics.
For instance, the reconstruction error in~\gls{ae}s can be adapted to incorporate survey weights as follows:
\[
\tilde{\mathcal{L}}_{\text{AE}}(\mathbf{x}; \theta, \varphi) 
= \sum_{j = 1}^{n} w_{j} \, \big\| \mathbf{x}_j - g_{\varphi}(f_{\theta}(\mathbf{x}_j)) \big\|^2,
\]  
where $w_j$ is the weight of the $j$-th sample and $f_{\theta}(\cdot)$ and $g_{\varphi}(\cdot)$ denote the encoder and decoder mappings, respectively. In this formulation, each unit contributes to the reconstruction error in proportion to its survey weight, ensuring that the model emphasises reconstruction fidelity in line with the target population.
Similarly, in~KPCA, the mapping $\Phi$ must be determined from a kernel matrix whose feature space representation has zero mean. When dealing with complex survey data, this centring step must account for the calibrated sampling weights $w_j$.  
Denoting by $\mathbf{K}_{ij}$ the entries of the kernel matrix, we obtain the desired centred kernel matrix $\tilde{\mathbf{K}}_{ij}$ (or Gram Matrix) by
\begin{equation*}
    \tilde{\mathbf{K}}_{ij} = \mathbf{K}_{ij} 
    - \sum_{l = 1}^{n} w_l\mathbf{K}_{il}
    -\sum_{l = 1}^{n} w_l\mathbf{K}_{lj} + 
    \sum_{m = 1}^{n}\sum_{l = 1}^{n} w_lw_m\mathbf{K}_{lm}.
\end{equation*}
For clarity and completeness, the formal notation and derivations are deferred to Appendix~\ref{apx: Incorporating_Sampling_Weights}.

\section{Results Application to EU-SILC Liguria data 2019: Empirical Results} \label{sec:Results}

This section presents the results of the application of the methodology introduced in Section~\ref{sec: methodology}, with the purpose of revealing potential layers of under-represented subgroups within the Ligurian subset of the 2019~EU-SILC data.
    
\subsection{Detection of misrepresented subgroups}

As introduced in Section~\ref{sec: methodology}, the analysis was framed as an outlier detection problem, for which we employed three distinct methodologies. 
For each method, we present both its effectiveness and the corresponding validation analyses.

The outlier detection outcomes derived from the Entropy Score method in Sub-section~\ref{sec: Entropy Score} are summarised in Table~\ref{tab: entropy_results}. 
Among the atypical variables which individually may indicate the presence of under-represented subgroups, we identified seven related to specific household characteristics. 
Specifically, these include whether household members are expected to reside in the dwelling for the entire reference year, the possibility of holding a second citizenship different from the Italian one, and the nature of each member’s relationship to Italian citizenship (including whether they possess it, how it was acquired, and whether they were born with it). 
Other relevant variables pertain to the type of school attended by household members (when applicable), and, more generally, the severity of material deprivation.
\begin{table}[]
    \centering
    \begin{tabular}{|l|c|c|}
    \toprule                                
    Selected~EU-SILC Variables  &  Entropy Score &  Cluster \\ 
    \midrule  
    Lived within the household for the whole 2019 &       0.951527 & Atypical \\           
    Secondary Citizenship (if applicable) &       0.874917 & Atypical \\                
    Severity of material deprivation &       0.804957 & Atypical \\         
    Type of school attended (if applicable) &       0.737139 & Atypical \\ 
    Italian Cizitenship &       0.725303 & Atypical \\  
    Italian citizen from birth &       0.714841 & Atypical \\
    Continuous residence in Italy &       0.657279 & Atypical \\  
    Risk of poverty indicator &       0.498129 &  Typical \\   
    Number of minor Children within the household &       0.459679 &  Typical \\
    Number of women aged 15-55 &       0.405058 &  Typical \\ 
    Social exclusion due to poverty &       0.375726 &  Typical \\ 
    Source of income or funding &       0.288991 &  Typical \\ 
    Civil status &       0.270583 &  Typical \\   
    Indicator of low work intensity &       0.241451 &  Typical \\ 
    Number of individuals in the household &       0.216393 &  Typical \\  
    Employment status or occupation type &       0.143095 &  Typical \\     
    Respondent's age &       0.127812 &  Typical \\  
    Type of family &       0.117092 &  Typical \\ 
    Quintile in the European Union economic ranking &       0.026564 &  Typical\\ 
    \bottomrule
    \end{tabular}
    \caption{Outlier variable detection results using the Entropy Score. The first column provides a description of the 19~EU-SILC variable under consideration. The central column reports the corresponding Entropy Score values. The third column presents the results of a 2-cluster K-means algorithm.}
    \label{tab: entropy_results}
\end{table}

For a more detailed investigation, we examined the sampling  (marginal) distribution of these variables, as shown in Table~\ref{tab:entropy_histograms}. 

Among the categories with low prevalence ($\le 10\%$), we observed several noteworthy subgroups: individuals who declared they do not hold Italian citizenship ($\sim 5\%$); those who spent a medium- to long-term period abroad (excluding tourism and other leisure-related travel) ($\sim 6\%$); individuals who are not Italian from birth ($\sim 5\%$); those who possess a second citizenship ($\sim 2\%$); and those experiencing material deprivation ($\sim 3\%$). 
Additional attention is drawn to school attendance ($\sim 8\%$), particularly to individuals with no schooling ($\sim 0.08\%$) or those enrolled in nursery or kindergarten ($\sim 0.04\%$).
Finally, we also identified households in which one or more members did not reside continuously throughout the entire reference year ($< 1\%$).

\begin{table}[htbp]
\centering
\renewcommand{\arraystretch}{1.2} 
\begin{tabularx}{\textwidth}{|l|X|}
\toprule
\textbf{Variable Name} & \textbf{Category Frequencies} \\
\midrule
Continuous residence in Italy & 
Yes: 0.9361 \\ 
& No: 0.0639 \\ 
& Not applicable: 0 \\[2pt]
\hline
Italian citizen from birth & 
Yes: 0.9257 \\ 
& No: 0.0473 \\ 
& Not applicable: 0.027 \\[2pt]
\hline
Secondary citizenship & 
Yes: 0.0171 \\ 
& No: 0.9829 \\ 
& Not applicable: 0 \\[2pt]
\hline
Severity of material deprivation & 
Yes: 0.0301 \\ 
& No: 0 \\ 
& Not applicable: 0.9699 \\[2pt]
\hline
Type of school attended & 
Kindergarten: 0.019 \\ 
& Nursery: 0.024 \\ 
& Primary school: 0.036 \\ 
& Lower secondary school: 0.017 \\ 
& Upper secondary school: 0.004 \\ 
& No schooling: 0.007 \\ 
& Not applicable: 0.893 \\[2pt]
\hline
Lived within the household for the whole 2019 & 
Yes: 0.992 \\ 
& No, only for a period: 0.008 \\ 
& No: 0 \\[2pt]
\hline
Italian citizenship & 
Yes: 0.954 \\ 
& No: 0.047 \\
\bottomrule
\end{tabularx}
\caption{The observed values (frequency) within the categorical variables identified as atypical and potentially indicative of misrepresented subgroups according to the Entropy Score. The left column lists the variable names and the right column reports the relative frequencies of each category.}
\label{tab:entropy_histograms}
\end{table}

For the stability validation of the {KPCA} and the {AE} methods, we obtained~MCC of $0.93 \pm 0.01$ and $0.94 \pm 0.01$, respectively.
Additionally, for the internal validation,~MCC were equal to $0.98 \pm 0.01$ and $0.97 \pm 0.01$, respectively.
For the variables inspection, we refer to Tables~\ref{tab: Importance_KPCA} and~\ref{tab: Importance_AE}.
According to the~{KPCA} in Table~\ref{tab: Importance_KPCA}, the top-5 variables that are most influential in outliers identification are (in decreasing order of importance)  the respondent's age, the number of minor children within the household, social exclusion due to poverty, low-work intensity, and the risk of severe material deprivation.
According to the~{AE} in Table~\ref{tab: Importance_AE}, the top-5 variables that are most influential in the outlier identification are, in decreasing order of importance,  the employment status, the main source of funding for the household, the respondent's age, the number of individuals in the households, and the low work intensity.
\begin{table}[]
    \centering
 \begin{tabular}{|l|c|} 
 \toprule                                    Selected~EU-SILC Variables &  Average Importance (KPCA) \\ 
 \midrule                                Respondent's age &                   0.728312 \\
 Number of minor children within the household &                   0.727142 \\                 Social exclusion due to poverty &                   0.726549 \\
 Indicator of low work intensity &                   0.725653 \\ 
 Risk of poverty indicator &                   0.724115 \\
 Severity of material deprivation &                   0.723654 \\
 Type of family &                   0.722987 \\  
 Civil status &                   0.721763 \\ 
 Number of women aged 15-55 &                   0.720985 \\          
 Number of individuals in the household &                   0.719878 \\ 
 Italian Citizenship &                   0.718312 \\ 
 Source of income or funding &                   0.717764 \\            
 Employment status or occupation type &                   0.715522 \\   
 Lived within the household for the whole 2019 &                   0.714221 \\         
 Type of school attended (if applicable) &                   0.713702 \\
 Never moved out of Italy &                   0.713145 \\           
 Secondary Citizenship (if applicable) &                   0.712773 \\ 
 Italian citizen from birth &                   0.711452 \\ 
 Quintile in the European Union economic ranking &                   0.708764 \\ 
 \bottomrule 
\end{tabular}
    \caption{Average Importance for variables in KPCA}
    \label{tab: Importance_KPCA}
\end{table}
\begin{table}[]
    \centering
    \begin{tabular}{|l|c|} 
    \toprule                                    Selected~EU-SILC Variables &  Average Importance (AE) \\ 
    \midrule            Employment status or occupation type &                 0.204641 \\                     Source of income or funding &                 0.176416 \\ 
    Respondent's age &                 0.115671 \\          
    Number of individuals in the household &                 0.110805 \\
    Indicator of low work intensity &                 0.080468 \\
    Type of school attended (if applicable) &                 0.076208 \\
    Civil status &                 0.054061 \\ 
    Number of minor Children within the household &                 0.049749 \\  
    Type of family &                 0.042771 \\ 
    Number of women aged 15-55 &                 0.016658 \\
    Lived within the household for the whole 2019 &                 0.005160 \\
    Quintile in the European Union economic ranking &                 0.004485 \\ 
    Italian Citizenship &                 0.001281 \\           
    Secondary Citizenship (if applicable) &                 0.000814 \\ 
    Italian citizen from birth &                 0.000762 \\    
    Risk of poverty indicator &                 0.000190 \\   
    Severity of material deprivation &                 0.000035 \\                       
    Continuous residence in Italy &                 0.000035 \\                
    Social exclusion due to poverty &                 0.000000 \\ 
    \bottomrule 
    \end{tabular}
    \caption{Average Importance for variables in AE}
    \label{tab: Importance_AE}
\end{table}

The medoids utilised to explore some profiles representing the potential under-represented groups are shown in Tables~\ref{tab: Medoids_KPCA} and~\ref{tab: Medoids_AE}
For the~\gls{kpca} analysis, we found four subgroups, which capture markedly different socio-demographic profiles, each reflecting unique household compositions and degrees of socio-economic vulnerability (see Table~\ref{tab: Medoids_KPCA}).

Subgroup 1 is composed of minors (aged 0-16) residing in households with three members, typically within nuclear families consisting of two parents and at least one child. 
These households show no signs of material deprivation or poverty, and the child attends an upper secondary school. 
The presence of one woman aged in the class 15-55 and one minor child reinforces the idea of a young, structurally stable family unit with regular schooling and no significant economic hardship.

Subgroup 2 represents older individuals (aged over 64), often widowed and living alone. These households consist of two members, possibly indicating the presence of a caregiver or companion, and are entirely supported by pensions. 
The profile reflects a typical post-retirement condition, where individuals have Italian citizenship from birth, and are not exposed to material deprivation or poverty. Their household structure and reliance on pensions suggest financial independence but also potential social isolation.

Subgroup 3 reveals a more fragile and marginalised segment of the population. 
It comprises individuals aged 16-32 in single-parent households with only "adult children". 
These households are flagged by poverty risk and indicators of social exclusion. 
Members in this group often have no secondary citizenship, are not Italian by birth, and have potentially experienced international migration. Despite being in the working-age bracket, employment is not the primary source of income, suggesting precarious economic conditions. This subgroup, while demographically younger, is marked by social vulnerability and limited structural support.

Subgroup 4 encompasses economically active adults aged 32-64, typically employed and part of dual-adult households without minor children. 
These households are sustained by salaried income and include one woman of reproductive age, with no minors and no signs of deprivation or poverty. 
The structure indicates a relatively stable socio-economic status, fitting into the middle-income bracket of the European economic ranking.
Although exposed to fewer risks than Subgroup 3, this profile may still reflect households in transition, possibly with grown-up children and a focus on economic consolidation.

In Table~\ref{tab: Medoids_AE} we distinguish seven subgroups.
Subgroups 1 and 2 are composed of young adults (aged 16-32), with Italian citizens by birth, living in medium-sized households (4 members), and relying primarily on support from cohabiting family members. 
While Subgroup 1 is economically disadvantaged and flagged by poverty and social exclusion, Subgroup 2 occupies a more stable economic position, ranking in the top EU income quintile.
Subgroup 3 contains older adults (over 64), living in large, extended households with multiple family units. 
This group is supported by pensions and shows no poverty indicators, suggesting a degree of economic stability within a multi-generational living arrangement.
Subgroup 4 includes working-age adults (32-64), typically divorced and employed, living in small households. 
They rely on salaried income and rank in the middle of the EU income distribution, indicating a profile of economic independence and stability.
Subgroup 5 stands out for its very large households (9 members), with several minor children and women of reproductive age. 
The individuals are married and have dual citizenship, with a continuous residence in Italy, and rely on family-based income support. 
This suggests a traditional and possibly multicultural family structure with complex dependencies.
Subgroup 6 reflects a smaller, nuclear family setup with minor children and one working parent, indicating a relatively stable but support-reliant structure.
Subgroup 7 portrays a mature, married couple without children, where the woman is aged 35 - 64, both employed and in the highest EU income quintile. This subgroup appears to be economically privileged and demographically stable.

\begin{longtable}{|p{4cm}|p{2cm}|p{2cm}|p{2cm}|p{2cm}|}
\toprule
 Selected~EU-SILC Variables & Subgroup 1 & Subgroup 2 & Subgroup 3 & Subgroup 4 \\
\midrule
\endfirsthead

\toprule
 Selected~EU-SILC Variables & Subgroup 1 & Subgroup 2 & Subgroup 3 & Subgroup 4 \\
\midrule
\endhead

Respondent's age & 0--16 & Over 64 & 16--32 & 32--64 \\ \hline

Italian Citizenship & YES & YES & YES & YES \\\hline

Italian citizen from birth & YES & YES & NO & YES \\ \hline

Secondary Citizenship (if applicable) & NO & NO & NO & NO \\ \hline

Continuous residence in Italy & Not applicable & Not applicable & Not applicable & Not applicable \\ \hline

Type of school attended (if applicable) & Upper secondary school & Not applicable & Not applicable & Not applicable \\ \hline

Lived within the household for the whole 2019 & YES & YES & YES & YES \\ \hline

Employment status or occupation type & Not applicable & Retired & Other & Employed \\ \hline

Source of income or funding & Not applicable & Pensions & Support from cohabiting family members & Income from salaried employment \\ \hline

Number of individuals in the household & 3 & 2 & 3 & 3 \\ \hline

Number of minor children within the household & 1 & 0 & 0 & 0 \\ \hline

Number of women aged 15--55 & 1 & 0 & 1 & 1 \\ \hline

Civil status & Single & Widowed & Single & Single \\ \hline

Type of family & Couple with at least one minor child & Single aged 65 and over & Single parent with only adult children & Couple with only adult children \\ \hline

Severity of material deprivation & Not applicable & Not applicable & Not applicable & Not applicable \\ \hline

Risk of poverty indicator & Not applicable & Not applicable & YES & Not applicable \\ \hline

Indicator of low work intensity & NO & Not applicable & NO & NO \\ \hline
Social exclusion due to poverty & NO & NO & YES & NO \\ \hline

Quintile in the European Union economic ranking & 3rd & 5th & 1st & 3rd \\
\bottomrule
\caption{Medoids for each subgroup based on the analysis thought~\gls{kpca} of~EU-SILC variables } 
\label{tab: Medoids_KPCA}
\end{longtable}

\begin{longtable}{|p{3cm}|p{1.5cm}|p{1.5cm}|p{1.5cm}|p{1.5cm}|p{1.5cm}| p{1.5cm}| p{1.6cm}|}
\toprule
 Selected~EU-SILC Variables & Subgroup 1 & Subgroup 2 & Subgroup 3 & Subgroup 4 &  Subgroup 5 & Subgroup 6 & Subgroup 7\\
\midrule
\endfirsthead

\midrule                                
Respondent's age &                                  16-32 &                                  16-32 &                      Over 64 &                           32-64 &                                  32-64 &                                  32-64 &                                      32-64 \\ \hline                             
Italian Citizenship &                                    YES &                                    YES &                          YES &                             YES &                                    YES &                                     NO &                                        YES \\ \hline                      
Italian citizen from birth &                                    YES &                                    YES &                          YES &                             YES &                                     NO &                         Not applicable &                                        YES \\ \hline    
Secondary Citizenship (if applicable) &                                     NO &                                     NO &                           NO &                              NO &                                    YES &                                     NO &                                         NO \\ \hline   
Continuous residence in Italy &                         Not applicable &                         Not applicable &               Not applicable &                  Not applicable &                                    YES &                                    YES &                             Not applicable \\ \hline         
Type of school attended (if applicable) &                         Not applicable &                         Not applicable &               Not applicable &                  Not applicable &                         Not applicable &                         Not applicable &                             Not applicable \\ \hline   
Lived within the household for the whole 2019 &                                    YES &                                    YES &                          YES &                             YES &                                    YES &                                    YES &                                        YES \\ \hline            
Employment status or occupation type &                               Employed &                               Employed &                      Retired &                        Employed &                                  Other &                               Employed &                                   Employed \\ \hline                     
Source of income or funding & Support from cohabiting family members & Support from cohabiting family members &                     Pensions & Income from salaried employment & Support from cohabiting family members & Support from cohabiting family members &     Support from cohabiting family members \\ \hline          
Number of individuals in the household &                          4 &                          4 &                6 &                   3 &                         9 &                          3 &                              2 \\ \hline   
Number of minor Children within the household &                            0 &                           0 &                  2&                     0 &                            3 &                            1&                                0\\ \hline
Number of women aged 15-55 &                            1 &                            1&                 1&                     1 &                            4 &                            1 &                                0\\ \hline     
Civil status &                                 Single &                                 Single & Married (living with spouse) &                        Divorced &           Married (living with spouse) &           Married (living with spouse) &               Married (living with spouse) \\ \hline         
Type of family &        Couple with only adult children &        Couple with only adult children &     Two or more family units &               Single aged 35--64 &               Two or more family units &   Couple with at least one minor child & Couple without children – Woman aged 35--64 \\ \hline  
Severity of material deprivation &                         Not applicable &                         Not applicable &               Not applicable &                  Not applicable &                         Not applicable &                         Not applicable &                             Not applicable \\ \hline  
Risk of poverty indicator &                                    YES &                         Not applicable &               Not applicable &                  Not applicable &                         Not applicable &                         Not applicable &                             Not applicable \\ \hline   
Indicator of low work intensity &                                     NO &                                     NO &               Not Applicable &                              NO &                                     NO &                                     NO &                             Not Applicable \\ \hline  
Social exclusion due to poverty &                                    YES &                                     NO &                           NO &                              NO &                                     NO &                                     NO &                                         NO \\ \hline 
Quintile in the European Union economic ranking &                                    1st &                                    5th &                          2nd &                             4th &                                    2nd &                                    4th &                                        5th \\ 
\bottomrule
\caption{Medoids for each subgroup based on the AE analysis of~EU-SILC variables } 
\label{tab: Medoids_AE}
\end{longtable}

\subsection{Models' validation}
According to Sections~\ref{sec: Stability_Validation} and~\ref{sec: Internal_Validation}, all implemented models were subjected to two complementary validation procedures. The \textit{stability validation} assessed the robustness of each model to the removal of individual instances, thus evaluating its sensitivity to single-case perturbations. The \textit{internal validation} examined the model’s ability to capture information that is consistent and transversal across the dataset. 
For the Entropy Score, the stability validation yielded an~MCC of $0.94$ (95\% CI: $0.92$--$0.96$), while the internal validation resulted in an~MCC of $0.91$ (95\% CI: $0.89$--$0.93$).  
For the~\gls{kpca}, we obtained an~MCC of $0.99$ (95\% CI: $0.97$--$1.00$) in the stability validation and $0.98$ (95\% CI: $0.97$--$0.99$) in the internal validation.  
Finally, for the~\gls{ae}, both the stability and internal validations achieved an~MCC of $0.99$ (95\% CI: $0.98$--$1.00$), confirming the high reliability and internal consistency of the empirical results presented above

\section{Integrative Sampling Strategies for Enhanced subgroup Representation}
\label{sec: Intergative_Sampling}

As outlined in Section~\ref{sec: intro}, the Ligurian~EU-SILC dataset serves as a case study to examine how data-driven methodologies can identify misrepresented subpopulations, i.e. groups that may correspond to frail, marginalised, or otherwise underserved communities requiring targeted attention from local policy frameworks. 
The proposed methodological approach  (Section~\ref{sec: methodology}) advances beyond conventional adjustment procedures by explicitly detecting and profiling misrepresented subgroups through accounting for the sample design and addressing distortions that emerge when vulnerable segments of the population are insufficiently captured. 
In line with the recommendations of~\cite[]{NASEM2023}, which call for all populations, particularly those historically underrepresented, to be made visible and accurately portrayed, the approach presented here integrates methodological innovation with an ethically informed, policy-oriented perspective without compromising analytical integrity.

Building upon the profiling of misrepresented subpopulations, the need to improve data on these groups becomes evident. This necessity was acknowledged in the Liguria Project (see Section~\ref{sec: liguria_project}), leading to the consideration of a supplementary sample survey specifically targeting the population strata identified as misrepresented and potentially vulnerable, those that social policy interventions should prioritise. In this regard, the enhancement of the Ligurian~EU-SILC dataset would primarily target households identified under the Resilience and Family Plan and those empirically determined to be underrepresented through linkage and comparison with Tax Declaration Data (see Section~\ref{sec: data_tax_decalration}). 
Accordingly, the present section explores possible designs for a tailored complementary survey focused on the subpopulations identified as underrepresented through the proposed data-driven methodology. A simulation exercise based on artificial data is presented to compare its applicability, relative strengths, and limitations. The purpose is twofold: first, to provide a methodological discussion; and second, to derive practical recommendations for supplementing EU-SILC data and promoting empirically grounded policymaking at the regional level.

\subsection{Simulated Sampling Strategies} 
\label{sec: Simulated_strategies}

Two sampling strategies were considered, each defined as a pair consisting of a sample design and an estimator for the population quantity of interest. These strategies differ in their data and contextual requirements for practical survey implementation. The rationale behind both is to control and tailor the sample size for specifically targeted subgroups, while accounting for varying levels of availability of both sampling frames and population-level para-data necessary for individual selection and data collection. 

The first strategy involves a stratified sample design, which can be adopted when two key data requirements are met:
\begin{enumerate}
    \item the availability of high-quality auxiliary data at the regional level, such as official data from a recent census, pertaining to one or more variables identified as crucial for detecting the targeted subgroups. Such auxiliary variables can be used as stratification variables to enhance the prevalence of individuals belonging to the misrepresented subgroups identified through the data-driven analysis, within one or more strata of the population, using an appropriately chosen allocation of the total fixed sample size; and
    \item the existence of a complete list of the regional population, including all subgroups of interest for the complementary data collection, which can serve as a reliable sampling frame for both stratification and  sample selection~\cite{barron2015using}.
\end{enumerate}
When these conditions are satisfied, standard unbiased estimation methods and sample size allocation procedures can be used to account for both budgetary and accuracy constraints. In practice, however, these requirements may be difficult to meet. A practical alternative is to use two or more partial sampling frames, either pre-existing or purposely constructed, expected to exhibit a significant prevalence of the identified target subgroups, the higher the better. This leads to the second strategy, a Multiple Frame Survey (MFS) involving a number $Q \ge 2$ of partial frames. An independent sample of pre-specified size is drawn from each frame, and the collected data are then combined to produce the desired estimates. An MFS can be implemented under the following conditions: 
\begin{enumerate}
    \item although individually partial, the union of the $Q$ frames can be trusted to adequately cover the targeted regional subgroups of interest for the complementary data collection; and 
    \item the $Q$ frames can, and often do, overlap, and the same subject may appear in more than one frame. The overlap may facilitate the construction of usable partial frames, but simultaneously introduces a statistical complication: units appearing in more than one frame have an increased probability of selection. This issue, known as \emph{multiplicity}, is accounted for through  the adoption of a Multiple Frame estimator that incorporates multiplicity-adjustement of the design weights. The computation of such adjustments requires additional data on units' frame membership, which must therefore be collected during the survey.
\end{enumerate}
From the extensive literature on Multiple Frame Surveys (see, for instance,~\cite[]{lohr2007recent, Lohr&Brick14}, two estimators are considered here: the \emph{Simple Multiplicity} (SM) estimator \citep{mecatti2007single,Mecatti&Singh14} and the \emph{Pseudo Maximum Likelihood} (PML) estimator~\citep[]{skinner1996estimation, lohr2006multiple}. These estimators differ in both estimation principles and data requirements for multiplicity adjustment. The SM estimator represents the simplest approach, requiring only basic information on sampled units' frame membership and allowing for an easy-to-compute fixed multiplicity correction. It is unbiased and operationally practical, though it may lack efficiency. The PML estimator is nearly unbiased and typically more efficient, but more complex to implement and requiring more detailed frame membership information. 
Further details see~\cite{mecatti2007single}. 
Finally, both strategies considered here encompass, as special cases, ad hoc integrative sampling campaigns specifically designed to target a single subgroup identified as underrepresented.

\subsection{Simulative Scenarios}
\label{sec:simulativescenarios}

\begin{table}[htbp]
\centering
\begin{tabular}{|c|c|c|c|c|}
\toprule
\multicolumn{5}{|l|}{\textbf{Population}} \\
\midrule
\multicolumn{5}{c}{
\begin{minipage}{0.9\textwidth}
\centering
\includegraphics[width=0.9\textwidth]{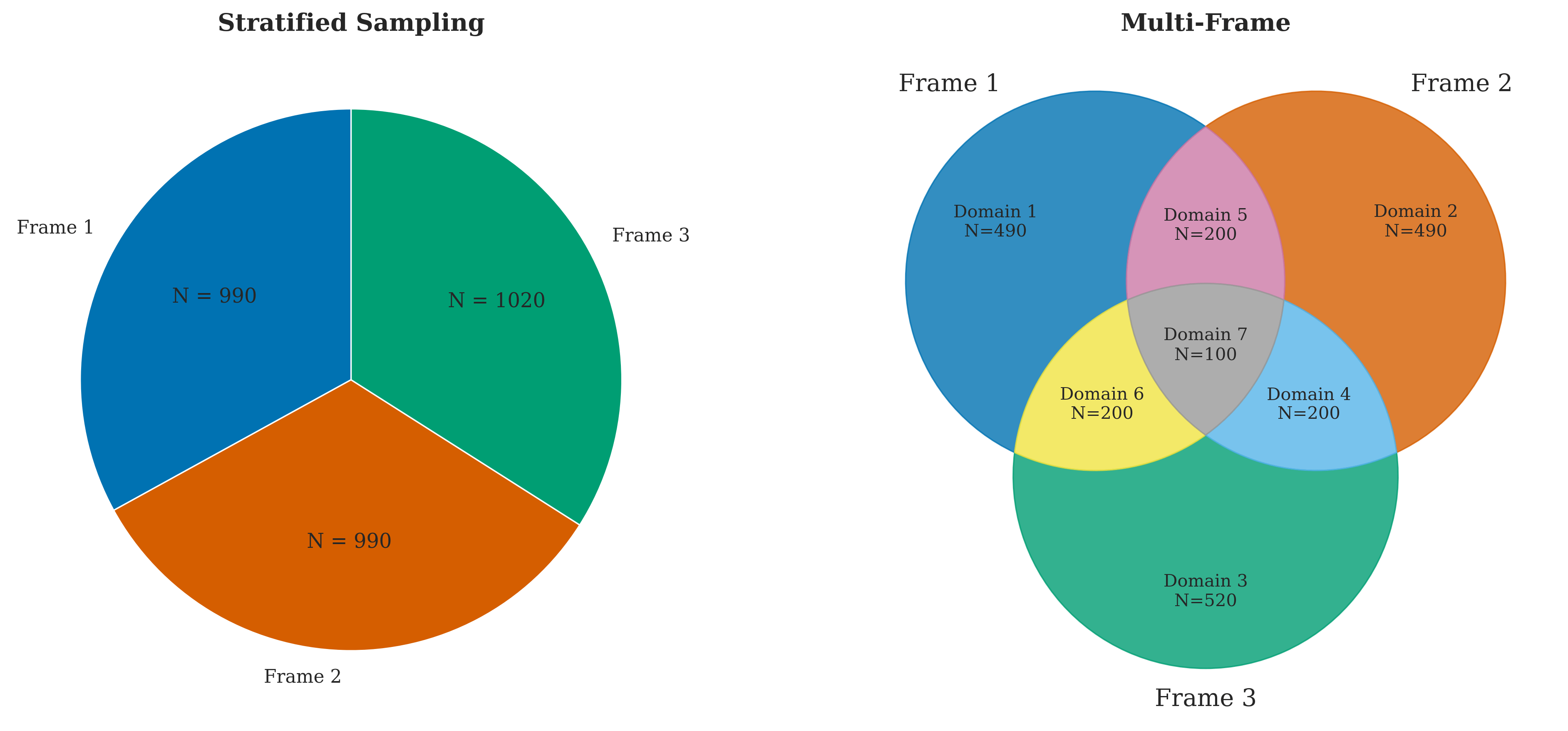}
\end{minipage}
} \\
\midrule
\multicolumn{5}{|l|}{\textbf{Sample}} \\
\midrule
\textbf{Design} & \textbf{Frame} & \textbf{Allocation} & \textbf{Sampling Fraction} & \textbf{Sample Size} \\
\midrule
  Stratified (STS) & Frame 1 & Proportional & 0.20 & 198 \\
  Stratified (STS) & Frame 2 & Proportional & 0.20 & 198 \\
  Stratified (STS) & Frame 3 & Proportional & 0.20 & 204 \\
\hline  Stratified (STS) & Frame 1 & Optimal Cost & 0.35 & 213 \\
  Stratified (STS) & Frame 2 & Optimal Cost & 0.28 & 171 \\
  Stratified (STS) & Frame 3 & Optimal Cost & 0.36 & 217 \\
\hline  Multi-Frame (MF) & Frame 1 & Proportional & 0.20 & 198 \\
  Multi-Frame (MF) & Frame 2 & Proportional & 0.20 & 198 \\
  Multi-Frame (MF) & Frame 3 & Proportional & 0.20 & 204 \\
\hline  Multi-Frame (MF) & Frame 1 & Optimal Cost & 0.26 & 156 \\
  Multi-Frame (MF) & Frame 2 & Optimal Cost & 0.41 & 246 \\
  Multi-Frame (MF) & Frame 3 & Optimal Cost & 0.33 & 198 \\
\bottomrule
\end{tabular}
\caption{Population and Sample Allocation under Stratified and Multi-Frame Sampling Designs}
\label{tab:allocation}
\end{table}

To implement the proposed sampling strategies and assess their relative effectiveness, a simulation study was conducted. A synthetic population of size $N = 3000$ was generated from a bivariate random variable $(X, Y)$ with a Gaussian distribution with parameters $\mu_x = 4$, $\mu_y = 1$, $\sigma_x = 0.3$, $\sigma_y = 0.2$, and $\rho_{xy} = 0.85$. 
In order to reflect naturally bounded socio-economic variables, the distribution was truncated at zero to ensure positivity of the population values for both variables. In this setting, $Y$ represents the study variable, with the population mean $\mu_y$ as the parameter to be estimated, and $X$ serves as the design variable.
For the stratified sampling strategy, the population was partitioned into $Q = 3$ strata based on the variable $X$. 
Specifically, three strata were constructed based on the 33rd and 66th percentiles of $X$, thereby producing approximately equal-sized strata in a practical and efficient manner.
while for the {MFS} strategy, the $N$ population units were randomly assigned to $Q = 3$ overlapping frames, with sizes reported in Table~\ref{tab:allocation}. For each strategy, the total sample size was fixed at $n = 600$ (corresponding to a sampling fraction of $n/N = 0.2$), and two sample allocation scenarios across the three strata or frames were considered: 
\begin{enumerate}
    \item a \emph{proportional allocation}, with the same sampling fraction ($0.2$) applied to each stratum or frame, and 
    \item an \emph{optimal allocation} \citep{Lohr&Brick14, Sarndal&al92, lohr2021sampling} under a cost constraint defined by a linear cost function:
    \[
    C = c_0 + \sum_{q=1}^Q n_q c_q,
    \]
    where $n_q$ denotes the sample size allocated to stratum (or frame) $q$. 
    The fixed cost was set to $c_0 = 0$, while the per-unit survey costs were set to $c_1 = 9.5$, $c_2 = 10$, and $c_3 = 10.5$.
\end{enumerate}
A simple random sample was selected within each stratum for the stratified sampling strategy (hereafter denoted as STS), whereas a Rao-Sampford sample design \cite[]{sampford1967sampling} was used within each frame for the MFS strategy. 
For each strategy and each allocation scenario, $M=2500$ samples were generated, producing simulated distributions of the three estimators for $\mu_y$ discussed in Subsection~\ref{sec: Simulated_strategies}. The relative performance of the sampling strategies was evaluated using standard Monte Carlo (MC)  metrics, namely the MC mean squared error (MSE) and  relative bias (RB) as given by:
\[
\operatorname{MSE}_{MC}(\hat{\mu}_y) = \frac{1}{M} \sum_{m=1}^{M} \left(\hat{\mu}_{y,m} - \mu_y\right)^2, \quad
\operatorname{RB}_{MC}(\hat{\mu}_{y}) = \frac{1}{M} \sum_{m=1}^{M} \frac{\hat{\mu}_{y,m} - \mu_y}{\mu_y}.
\]
where $\hat{\mu}_{y,m}$ denotes the estimate obtained at the $m$-th Monte Carlo iteration for the estimator under consideration, namely STS for the stratified design, and SM and PML for the MFS strategy. 
Further details on the simulation setup and the sample allocation procedures are provided in Appendix~\ref{apx:simulation_allocation}.

\subsection{Monte Carlo Results}
\begin{figure}[]
    \centering
    \begin{subfigure}{0.5\textwidth}
        \centering
        \includegraphics[width=\linewidth]{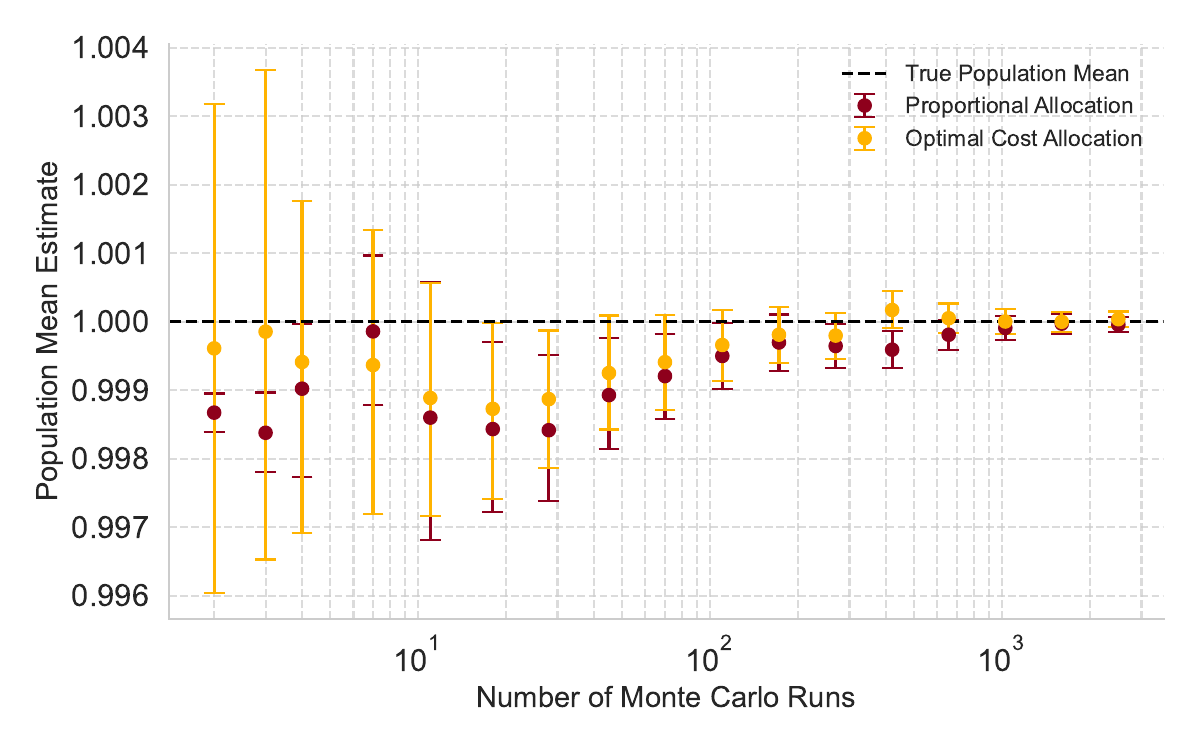}
        \caption{Stratified Sampling Design}
    \end{subfigure}
    \hfill
    \begin{subfigure}{0.5\textwidth}
        \centering
        \includegraphics[width=\linewidth]{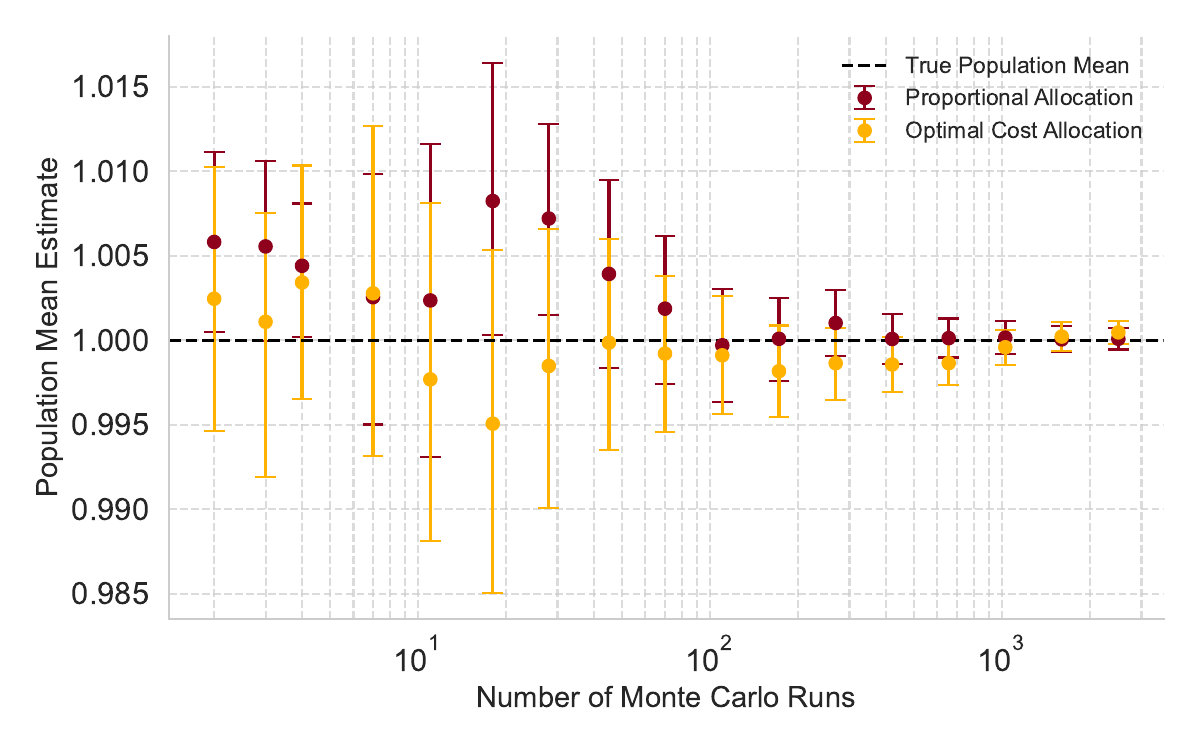}
        \caption{SM Estimator}
    \end{subfigure}
    \hfill
    \begin{subfigure}{0.5\textwidth}
        \centering
        \includegraphics[width=\linewidth]{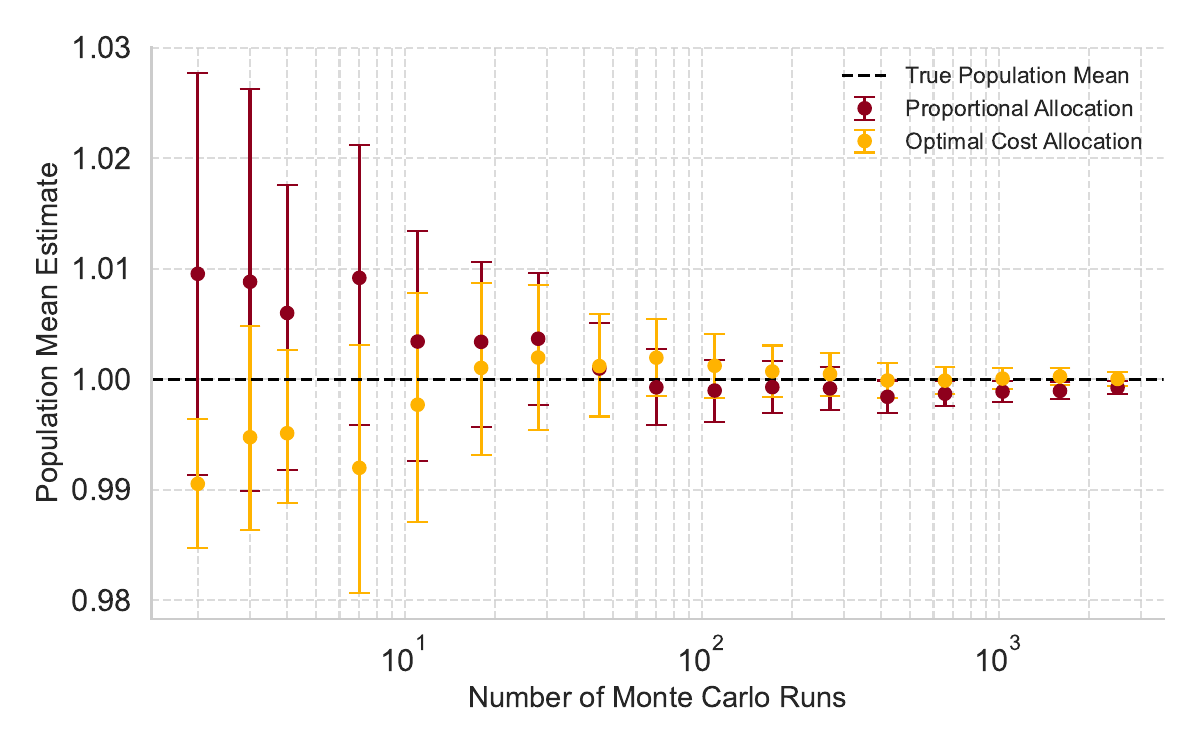}
        \caption{PML Estimator}
    \end{subfigure}
    
    \caption{Estimated population means with increasing numbers of Monte Carlo replications under three sampling scenarios: 
(a) stratified sampling design, 
(b) MF sampling design with the SM estimator, and 
(c) MF sampling design with the PML estimator. 
The trajectories illustrate the convergence of the estimators toward the theoretical population mean (dotted line). 
For each scenario, results are reported under both proportional and optimal cost allocation strategies. 
Error bars denote 95\% confidence intervals obtained via bootstrap resampling.}
    \label{fig: population_mean_estimate}
\end{figure}

To first evaluate the MC experiments, Figure~\ref{fig: population_mean_estimate} illustrates the convergence of the simulated estimator  (\gls{sts},~\gls{sm}, and~\gls{pml}) across iterations in the MC experiment. 
%
Based on these results, 2500 Monte Carlo iterations are  sufficient for the 
simulation setting considered, as at this point the interval defined by $P_5$ and $P_{95}$, i.e. the 5th and 95th percentiles of the MC empirical distributions, contains the true population mean, and the relative error between the estimated Monte Carlo mean and the true mean is effectively below 1\%. 
Simulation results are summarised in Table~\ref{tab: allocation_mse} according to the MC metric~$\operatorname{MSE}_{MC}$, across sampling strategies and sample size allocation scenarios.
Results are consistent with the data and context requirements discussed in Subsection~\ref{sec: Simulated_strategies}  for the actual implementation of the two sampling strategies simulated.
\begin{table}[h!]
\centering
\begin{tabular}{|c| c| c| c| c|}
\toprule
Allocation & Estimator &~\gls{mse} & $P_5$ & $P_{95}$ \\
\midrule
optimal\_cost & PML & $26.5 \times 10^{-6}$ & $23.3 \times 10^{-6}$ & $29.8 \times 10^{-6}$\\
optimal\_cost & SM  & $27.1 \times 10^{-6}$ & $23.8 \times 10^{-6}$ & $30.4 \times 10^{-6}$\\
optimal\_cost  & STS & $24.9 \times 10^{-6}$ & $21.6 \times 10^{-6}$ & $28.2 \times 10^{-6}$\\
\hline
proportional  & PML & $26.7 \times 10^{-6}$ & $23.4 \times 10^{-6}$ & $30.1\times 10^{-6}$\\
proportional  & SM  & $27.5 \times 10^{-6}$ & $24.1 \times 10^{-6}$ & $30.6 \times 10^{-6}$ \\
proportional  & STS & $26.2 \times 10^{-6}$ & $22.9 \times 10^{-6}$  & $27.5 \times 10^{-6}$ \\
\bottomrule
\end{tabular}
\caption{Comparison of average~\gls{mse} and confidence intervals for different allocation strategies and estimators.}
\label{tab: allocation_mse}
\end{table}

For the \gls{sts}, a slight advantage is observed for the optimal-cost allocation, with a relative difference of approximately 5\% compared to proportional allocation. 
In the multi-frame setting, the \gls{mse} is substantially lower when using the \gls{pml} estimator under both proportional and optimal-cost allocation, indicating a clear efficiency advantage.

\begin{figure}
    \centering
    \begin{subfigure}{0.48\textwidth}
        \centering
        \includegraphics[width=\linewidth]{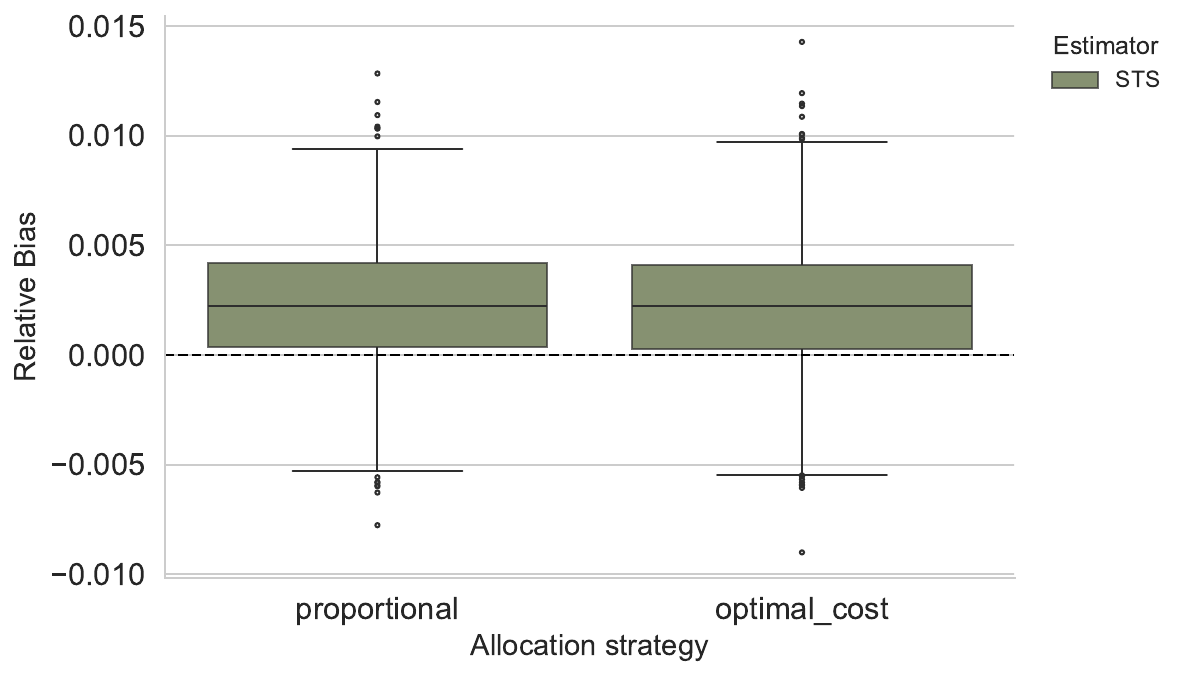}
        \caption{STS}
    \end{subfigure}
    \hfill
    \begin{subfigure}{0.48\textwidth}
        \centering
        \includegraphics[width=\linewidth]{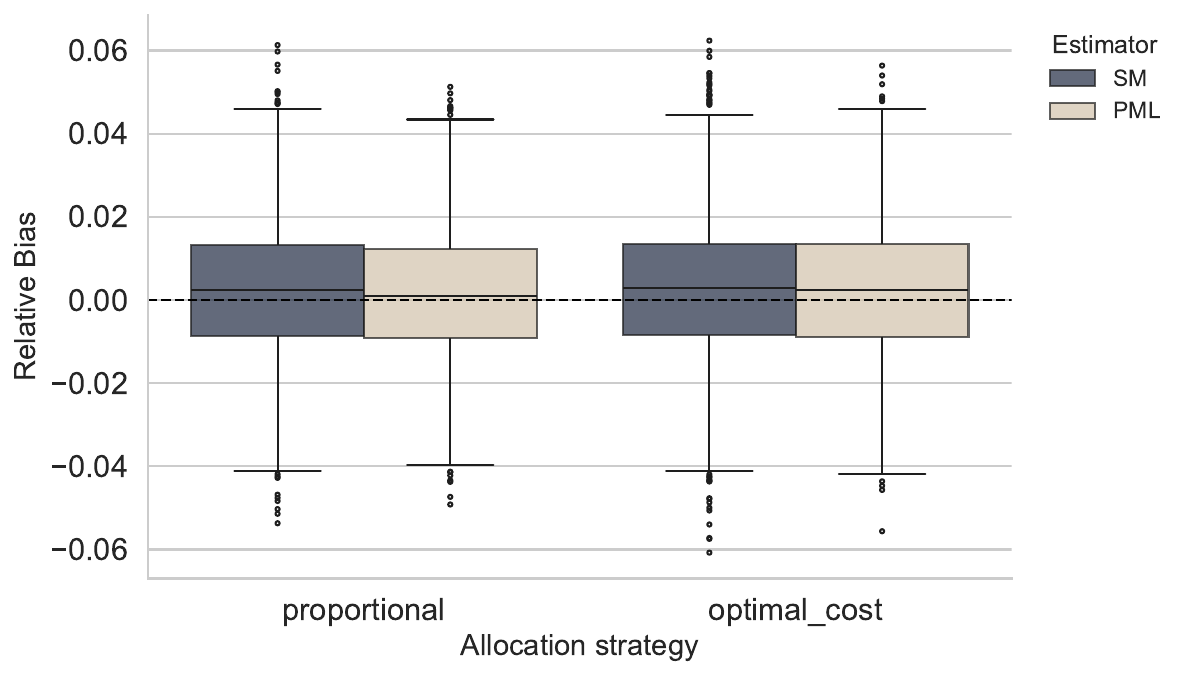}
        \caption{MF}
    \end{subfigure}
    \caption{Box plot for the RB estimation for STS and MF estimators across Monte Carlo iterations per allocation strategy}
    \label{fig: RB_simulation}
\end{figure}

Regarding the other MC metric~$\operatorname{RB}_{MC}$, in Figure~\ref{fig: RB_simulation} no substantial differences are observed across allocation strategies, with all boxplots centred around the ideal value of zero, consistently with the theoretical unbiasedness of \gls{sts} and \gls{sm} and quasi-unbiasedness of~\gls{pml}. 
In the~\gls{sts} case (Figure~\ref{fig: RB_simulation}a), the~\gls{rb} remains low, generally between -0.01 and 0.01, with a slight tendency for the median to lie above zero. 
In the multi-frame setting, the distributions are more symmetrically centred around zero, though with larger variability, ranging approximately from -0.05 to 0.05. 
Both the~\gls{sm} and~\gls{pml} estimators remain centred around zero and exhibit lower dispersion, reinforcing their stability.

These results emphasise that the choice of sampling strategy should be guided by the specific study context. 
When a complete sampling frame is available and stratification is feasible, a stratified sampling strategy provides a straightforward and practical option. 
In contrast, when only partial frames are accessible, particularly those containing characteristics of subgroups targeted for oversampling, an MFS strategy is recommended. 
Among the multi-frame estimators, the~\gls{pml} consistently demonstrates superior efficiency, although its higher para-data requirements and computational demands may limit practical applicability. 
In such cases, the~\gls{sm} estimator represents a viable alternative, combining 
theoretical unbiasedness with ease of para-data requirement and implementation. 
Moreover, multi-frame designs naturally accommodate differential frame costs: by accounting for the fact that some frames may be more expensive than others, cost-sensitive allocation can yield efficiency gains over standard unique frame approaches, such as the stratified strategy


\section{Discussion and conclusion}\label{sec: discussion}

As mentioned in Section~\ref{sec: intro}, 
the main objective of this work is to propose a statistical analysis of survey datasets, such as the~EU-SILC cross-sectional sample, aimed at identifying under-represented subgroups within the surveyed population. 
Although the~EU-SILC sampling design guarantees unbiased estimation of population parameters through  proper use of design weights, and allows valid inference at both national and sub-national level,~EU-SILC data may not provide sufficient information  for specific population subgroups possibly misrepresented in the sample.
Such imbalances may  pose significant practical issues, limiting or even potentially undermining the reliability of the dataset when used by public institutions  for national or regional socio-demographic policies and programs targeting under-served population subgroups.

Misrepresented subgroups often hide hard-to-reach or rare populations.
Their investigation requires prior knowledge of the subpopulations of interest, along with a sufficiently complete dataset that can serve as ground truth to reliably identify low-prevalence groups. 
Similarly, detecting under-coverage involves comparing the surveyed sample to a comprehensive reference dataset to identify which groups have been disproportionately represented. 
In both cases, the challenge lies in the unavailability of such a complete and reliable external dataset, especially in the context of socio-demographic studies.
Given this limitation, we reformulated the problem in terms of  outlier detection. 
By assuming that atypical patterns in the data may reflect either hard-to-reach groups or under-covered subpopulations, we employed a combination of traditional statistical methods and machine learning techniques to uncover these anomalies. 
The idea is that groups significantly deviating from the "core" characteristics of the surveyed population may be indicative of inadequate sampling representation.
In this framework, hard-to-reach populations, due to their inherent difficulty in being surveyed, tend to exhibit atypical patterns and thus naturally emerge as outliers. 
Similarly, groups that result misrepresented in the sample also appear as deviations from the bulk of the data.

Our approach included both univariate and multivariate models, aiming to explore multiple perspectives and validate findings across methods. 
The convergence of insights from diverse techniques constitutes a core strength of our approach. 
It not only allows for a more robust detection of under-represented subgroups but also enhances the interpretability and potential applicability of the results.
Our findings reveal specific household types that are misrepresented in the~EU-SILC data for Liguria, with a particular focus on complex and potentially vulnerable profiles. 
The entropy-based analysis identified critical gaps in the data, especially regarding households where individuals do not reside continuously, possess non-Italian citizenship (either primary or secondary) and lodge school-age children. Material deprivation also emerged as a key indicator of vulnerability of misrepresentation.
Multivariate analyses, particularly the~\gls{ae}, confirmed and deepened these insights.
It detected fragile groups, such as economically dependent youth at risk of poverty (e.g., Subgroup 1), and contrasted them with affluent both young and older individuals or stable, medium-sized households (e.g., Subgroups 2, 4, 6, 7 in Table~\ref{tab: Medoids_AE}).
A prominent finding is the emergence of large, multi-family, and multicultural households, often including three or more school-age children, non-Italian-born parents, and members with experience living outside Italy.  These households tend to have moderate or modest levels of wealth and are often overlooked by standard survey mechanisms (e.g., Subgroup 3).
The KPCA analysis, although less sensitive to foreign citizenship, confirmed patterns of vulnerability and affluence. 
Overall, the study reveals underrepresented subgroups along with distinct socio-economic contrasts within the Ligurian population.

The possibility of expanding population coverage by integrating~EU-SILC local sample to provide a more detailed and inclusive representation of regional and national subpopulations across Europe is implicit in our findings. 
Our methodology lays the groundwork for a user-driven complementary data collection, or data integration,
enabling targeted investigation of specific subpopulations through purpose-oriented sampling schemes. 
Beyond the~EU-SILC context, this methodology can be readily adapted to any similar scenario in which a surveyed population, suspected by the user to be misrepresented in available official statistics sample data, with respect to specific subgroups, requires data enrichment, using existing information as a foundation for more inclusive data collection.
Furthermore, the investigation into strategies aimed at addressing the integration of misrepresented subgroups through a simulation exercise based on two sampling strategies selected and discussed for practical application contexts,   indicates that no single method can be recommended as an all-purpose option, as all approaches exhibit context-dependent applicability and performance. 
The comparative study conducted in a MFS setting between the~\gls{sm} and~\gls{pml} estimators confirmed the superior efficency of the latter,  to be traded off with the ease of implementation and lesser para-data requirements of the former.
Accordingly, when a unique complete sampling frame is available and stratification is feasible, the~\gls{sts} approach is recommended due to its conceptual simplicity and methodological robustness. 
Conversely, in cases where only partial frames are accessible, a multi-frame strategy in conjunction with the~\gls{pml} estimator constitutes the preferred methodology.

Due to the absence of a reference or ground truth, all analyses were conducted in an unsupervised setting. 
This implies that the findings are inherently data-specific and lack external validation, which is typically provided in supervised learning contexts. 
Nevertheless, the internal validation and sensitivity analyses we performed offer a proof of concept, demonstrating that the methods tend to consistently capture similar features across different runs.
A potential limitation lies in the apparent dependence of the results with respect to the results. 
This study compared three different unsupervised approaches, each retrieving a set of overlapping findings while also focusing on distinct specifications. 
Although there is no universally agreed number of methods to employ, integrating multiple approaches may be advisable, following the practice in other research fields where ensemble or comparative methodologies are more common.

\section*{Acknowledgements}
The authors acknowledge the Bureaux of Financial Programming and the Statistical Study of Regione Liguria for the financial and expert support provided during the elaboration of this work. 
ER acknowledges the financial support from the MIUR Excellence Department Project 2023-2027 awarded to the Dipartimento di Matematica of the University of Genova, CUP D33C23001110001 and from the “Hub Life Science - Digital Health (LSH-DH) PNC-E3-2022-23683267-Progetto DHEAL-COM”, granted by the Italian Ministry of Health as part of the “PNRR Ecosistema Innovativo della Salute”.

\bibliographystyle{plainnat}  
\bibliography{references}  

\begin{appendices}
\noappendicestocpagenum
\addappheadtotoc

\section{Selected EU-SILC Variables}\label{apx: additional_info_data}

\begin{longtable}{ | m{4cm} | m{12cm}|}
    \toprule 
    \textbf{EU-SILC Acronym} &  \textbf{Variable's Description} \\
    
    \midrule
    ETAINT &  Age of respondent (categorised into: \st{0-1, 1-2, 2-4, 4-8, 8-16,} 16-32, 32-64, 65+ years) \\ \hline
    CITTADX & Italian Citizenship \\ \hline
    NCITT &  Italian citizen from birth \\ \hline
    SECITT &  Secondary Citizenship (if applicable) \\ \hline
    ITA & Continuous residence in Italy \\ \hline
    VIFAM &   Lived within the household for the whole 2019 \\ \hline
    LAVPRI & Employment status \st{or occupation type}\\ \hline  
    FONTERED & Source of income or funding  \\ \hline
    TOT\_TUTTI & Number of individuals in the household \\ \hline  
    RAGA017 & Number of minor Children within the household \\ \hline 
    DON1555 & Number of women aged 15-55 \\ \hline
    STACIV & Civil status \\ \hline 
    TF & Type of family \\ \hline
    TIPSCU & Type of school attended (if applicable) \\ \hline
    SEV\_MAT\_DEPRIV & Severity of material deprivation \\ \hline
    RISKPOV & Risk of poverty indicator \\ \hline
    LOW\_WORK\_INT & Indicator of low work intensity \\ \hline
    POV\_SOC\_EXCL & Social exclusion due to poverty \\ \hline 
    QUINTI\_EU &  Quintile in the European Union economic ranking \\ \bottomrule
    \caption{List with the description of the 19 categorical variables selected for the analysis. }
    \label{tab: descitpion eu-silc_data}
\end{longtable}

Table~\ref{tab: descitpion eu-silc_data} shows a summarising description of the variables extracted from the~EU-SILC dataset and their description. 
The \textit{severe material deprivation} (EV\_MAT\_DEPRIV) variable reflects the condition in which a person cannot afford at least four out of nine essential items. 
Let \texttt{N\_ITEM} be the number of items a person cannot afford, with values ranging from 0 to 9. 
Those items are  
\begin{enumerate}
    \item to pay their rent, mortgage or utility bills;
    \item to keep their home adequately warm;
    \item to face unexpected expenses;
    \item to eat meat or proteins regularly;
    \item to go on holiday;
    \item a television set;
    \item a washing machine;
    \item a car;
    \item a telephone.
\end{enumerate}
The severe material deprivation rate, broken down by subgroup $k$ (denoted \texttt{SEV\_DEPR\_TOTALat\_k}), is calculated using the adjusted weight \texttt{RB050a} as:
\[
\texttt{SEV\_DEPR\_TOTALat\_k} = 
\frac{\sum_{i \in k,\, N\_ITEM \geq 4} \texttt{RB050a}_i}
{\sum_{i \in k} \texttt{RB050a}_i} \times 100.
\]
With regard to the calculation of the above indicator, the threshold of four items to depict severe material deprivation has been chosen for a mixture of empirical and practical reasons since a previous threshold of 3 items had resulted in excessively high, and politically unmanageable, estimates of levels of deprivation across the EU. This gives the percentage of individuals in group $k$ experiencing severe deprivation; see~\cite[]{2024SevMatDepriv}.

The \textit{at-risk-of-poverty rate} (RISKPOV) indicates the proportion of individuals whose disposable income, adjusted for household size (equivalised), falls below 60\% of the national median after accounting for social transfers. 
This measure reflects income inequality rather than absolute poverty, highlighting how someone fares economically relative to the rest of the population in their country.
That is, this indicator does not necessarily imply poor living conditions but rather signals a lower income compared to the general population; see~\cite{2024Atriskpoverty}.

The indicator \textit{people living in households with very low work intensity}  (LOW\_WORK\_INT) captures the share of individuals aged 0--64 who live in households where the working-age members (18--64 years old) have, on average, worked less than 20\% of their full work-time potential during the previous year.
Work intensity is calculated as the ratio between the number of months actually worked by all working-age adults in the household and the total number of months they could theoretically have worked. These months are counted in full-time equivalent terms: part-time work is adjusted based on the number of usual working hours reported during the interview.
To ensure consistency, the calculation excludes individuals who are students aged 18--24, retirees (based on their self-declared economic status or pension receipt), and people aged 60--64 who are inactive and live in pension-dependent households. Households made up only of children, students under 25, or elderly people (65+) are also excluded from this indicator; see~\cite[]{2024LowWorkIntensity}

The indicator \textit{At Risk of Poverty or Social Exclusion} (AROPE) measures the share of the population facing at least one of the following three conditions: risk of poverty, severe material and social deprivation, or living in a household with very low work intensity (from the variable POV\_SOC\_EXCL). 
Each person is counted only once, even if they fall into multiple categories.
AROPE is the main metric used to track progress toward the European Union’s 2030 poverty and social exclusion targets, and it also serves as the headline indicator for the EU 2020 Strategy~\cite[]{2024AROPE}.

\section{Mathematical Details on Anomaly Detection Methods}\label{apx: dettagli_matematici_metodi_anomaly_detection}

\subsection{Entropy Score}\label{apx: Entropy_Score}

The \emph{Shannon entropy} (or simply, entropy) of a discrete random variable $X$, assuming values over a sample space $\mathcal X$ and distributed according to a probability mass function \( P \), is defined as 
\begin{equation}\label{eq:Entropy_discrete}
    H[P] = -\mathbf{E}_{X \sim P}[\log P(X)].
\end{equation}

By construction, \( H[P] \ge 0 \) and is a concave functional~\cite[]{bishop2006pattern}.  
For a discrete distribution with \( N \) categories, entropy is bounded between 0 and \( \log N \). The maximum value \( \log N \) occurs when all categories are equally likely, while the minimum value of zero corresponds to a degenerate distribution where a single category holds all the probability mass.

We used entropy in Section~\ref{sec: Entropy Score} as a scoring criterion to build an outlier model for identifying the most anomalous variables in the \gls{eusilc} dataset.
In the one-dimensional discrete setting, increasing entropy  reflects a progressive \emph{spreading} of the probability mass across categories. Thus, 
higher entropy corresponds to a more uniform distribution, whereas lower entropy signals a concentration of probability on a few categories.  This relationship between entropy and distributional spread is central to our sampling perspective.

Entropy and other information theory-based measures have long been used in anomaly and outlier detection~\cite[]{chandola2009anomaly}. 
Due to entropy's ability to quantify the informativeness of a source, it has also been proposed as a mechanism for automated selection of anomalous samples~\cite{daneshpazhouh2014entropy, berezinski2015entropy}.
In this work, we adapted entropy's properties to a univariate analysis framework, under the simplifying assumption that the dataset variables can be considered temporarily independent. 
Such a simplification allows the capture \emph{independent patterns} that most strongly characterise underrepresented subgroups. 
The method we propose is twofold: first, we consider a variable-specific entropy-based score to identify the most informative variables, those most likely to present outliers; second, we exploit the notion of \emph{information} associated with entropy to determine which categories within those variables are most representative of the target outliers.

Formally, let  \( V \)  be a univariate categorical variable taking values in the set of categories \( \{C_1, \dots, C_N\} \), with observed data \( x_1, \dots, x_n \).  
The entropy-based score we propose is the following:
\begin{equation}
    \Gamma_1 = 1+\frac{1}{\log N} \sum_{k=1}^{N} p_k \log p_k,
\end{equation}
where \( p_k \) denotes the empirical probability of category \( C_k \), given by:
\begin{equation}
    p_k = \frac{1}{n} \sum_{j=1}^{n} \mathbf{1}(x_j \in C_k).
\end{equation}
$\Gamma_1$ may take values only in $[0, 1]$; the higher the value of $\Gamma_1$, the more  the possibility of finding under-represented categories. 
Once the values of $\Gamma_1$ have been analysed to identify the top five most informative variables (see Section~\ref{sec: Outlier_Detection}), we examine the information content of each variable, quantified as $-\log p_k$, to determine which group is underrepresented within each selected variable.

\subsection{KPCA}\label{apx: KPCA}
Let $\hat{X} \in \mathbb{R}^{n \times p}$ be the data matrix  where $n$ is the number of sampled units and $p$ the number of variables. 
\gls{kpca} assumes the existence of a feature map $\Phi: \mathcal{X} \rightarrow \mathcal{H}$ into a Hilbert space $\mathcal{H}$, such that the inner product between any two mapped observations can be computed through a positive semidefinite kernel function: 
\[
k(\mathbf{x}, \mathbf{y}) = \Phi^{T}(\mathbf{x})  \Phi(\mathbf{y}).
\]
Therefore, \gls{kpca} treats the data as if they are mapped into a richer space where their structure becomes easier to separate. 
The kernel function enables the computation of how similar two unites are in this space without explicitly performing the mapping.

The transformed observations \( \Phi(\mathbf{x})\) are never computed explicitly.  
Instead, the analysis is carried out through the \emph{Gram matrix} \( \mathbf{K} \in \mathbb{R}^{n \times n} \), whose entries are defined as
\[
K_{xy} = k(\mathbf{x}, \mathbf{y}).
\]
Since \gls{pca} requires the data to be centred, the same operation must be performed in the feature space induced by \( \Phi \).  
Because this feature space is not explicitly available, centring is achieved by centring the Gram matrix itself.  
This operation produces the same effect as subtracting the mean of the mapped data in the (implicit) feature space.
In essence, centring the Gram matrix ensures that \gls{kpca} is conducted on mean-centred data, even though the feature map \( \Phi \) is never computed directly.

After centering $\mathbf{K}$ to obtain the \emph{centered Gram matrix} $\tilde{\mathbf{K}}$,~\gls{kpca} performs eigen-decomposition by solving:
\[
\tilde{\mathbf{K}} \mathbf{v} = \lambda \mathbf{v},
\]
where $\mathbf{v}$ are the eigenvectors corresponding to the principal directions in the feature space, and $\lambda$ are the associated eigenvalues.
In essence,~\gls{kpca} applies~\gls{pca} to the non-linearly transformed data $\Phi(\mathbf{x})$, enabling the capture of complex, non-linear structures in the original space. 
Unlike standard~\gls{pca}, the eigenvalues $\lambda$ no longer represent variances in the original input space, but instead quantify the variance of the data as expressed in the embedding space $\mathcal{H}$.

For outlier detection, an early application of~\gls{kpca} explored the geometry of the transformed feature space induced by the \emph{radial kernel}, as proposed in~\cite{hoffmann2007kernel}. 
In this framework, \emph{typical observations} are expected to lie close to the principal subspace 
$\{ \mathbf{v}_1, \dots, \mathbf{v}_k \}$, spanned by the top $k$ eigenvectors whose associated eigenvalues explain a significant portion (we opted for 99\%) of the variance in the embedding space $\mathcal{H}$. 

Given a new input $\mathbf{z} \in \mathbb{R}^{p}$, its image under the feature map is $\Phi(\mathbf{z}) \in \mathcal{H}$. 
The projection of $\Phi(\mathbf{z})$ onto the subspace defined by the top $k$ principal components is then:
\[
\hat{\Phi}(\mathbf{z}) = \sum_{i=1}^{k} [\mathbf{v}^{T}_i \Phi(\mathbf{z} )]\mathbf{v}_i,
\]
and the \emph{reconstruction error}, which serves as an unsupervised outlier score, is defined as:
\[
\operatorname{RE}_{\text{KPCA}}(\Phi(\mathbf{z})) = \| \Phi(\mathbf{z}) - \hat{\Phi}(\mathbf{z}) \|^2 = \sum_{i = k+1}^{p} \left| \mathbf{v}^{T}_i \Phi(\mathbf{z} ) \right|^2.
\]

To apply~\gls{kpca} in our context, which involves ordinal and categorical variables, we adopt a kernel based on the \emph{Hamming distance}.
We recall that the Hamming distance is given by:
\[
\operatorname{Ham}(\mathbf{x}, \mathbf{y}) = \sum_{i= 1}^{p}\mathbf{1}\{\mathbf{x}_i \neq \mathbf{y}_i\}. 
\]
Thus, the kernel we propose is:
\[
k(\mathbf{x}, \mathbf{y}) = \exp\left( -\gamma \cdot \operatorname{Ham}(\mathbf{x}, \mathbf{y}) \right),
\]
where $\gamma > 0$ is a bandwidth parameter controlling the rate of decay; for convenience, we set $\gamma = 1$.

This kernel function is positive semidefinite, a property required to ensure that the kernel defines a valid inner product in $\mathcal{H}$; see~\cite{shawe2004kernel}
More generally, we recall that metric distances, such as the Hamming or the Euclidean distance, can be transformed into valid kernels by applying the exponential function to distances themselves with the sign changed; see ~\cite{schoenberg1938metric, bekka2008kazhdan}.

\subsection{Integration of Design Weights in Anomaly Detection}\label{apx: Incorporating_Sampling_Weights}

In Section \ref{sec: Incorporating Sample Design Weights} we discussed how to  generalise any of the anomaly detection methods, and include sample weights in model training phases.
In practice, the weights can be directly applied to the training data and incorporated into the loss function. 
Thus, the model learns patterns that are representative of the target population, rather than being biased by the sample design.
%
For a generic variable \(X\), the weighted mean over a sample of size \(n\) is simply
\begin{equation}\label{eq: adjusted_average_two_stage_sampling}
\bar{x} = \sum_{j=1}^{n} w_j x_j,
\end{equation}
%
For instance, in the case of the \emph{Entropy Score}, which relies on histogram-based estimates, frequencies must be weighted accordingly. The weighted formulation becomes:
\[
\begin{cases}
    \tilde{\Gamma}_1 = -\frac{1}{\log N} \sum_{k=1}^{N} \tilde{p}_k \log \tilde{p}_k,\\
    \tilde{p}_k = \frac{1}{n} \sum_{j=1}^{n} w_{j} \mathbf{1}(x_j \in C_k).
\end{cases}
\]
where $w_{j} $ is the weight of the $j$ sampled unit.

For~KPCA, the eigen-decomposition is carried out on the centred Gram matrix 
$\tilde{\mathbf{K}}$ obtained once the contribution of the mean feature is removed in the embedding space $\mathcal{H}$, that is, 
\[
    \tilde{\mathbf{K}}_{xy} 
    \;=\; \tilde{\Phi}^{T}(\mathbf{x})\,\tilde{\Phi}(\mathbf{y}), 
    \qquad \text{with} \qquad \sum_{\mathbf{x}} \tilde{\Phi}(\mathbf{x}) = 0,
\]
where the centred feature map is defined as
\[
    \tilde{\Phi}(\mathbf{x}) \;=\; \Phi(\mathbf{x}) - \mu_{\mathbf{x}}.
\]
where $\mu_{\mathbf{x}}$ denotes the mean effect of the feature $\mathbf{x}$ 
under the distribution of samples. In the weighted setting, this mean depends 
on both the feature $\mathbf{x}$ and the sample weights $\mathbf{w} = 
(w_1,\dots,w_n)^{\top}$, such that
\[
    \mu \;=\; \sum_{j=1}^{n} w_j\,\Phi(\mathbf{x}_j), 
    \qquad
\]
where $\mathbf{w}$ are the same normalised weights used in Equation~(\ref{eq: adjusted_average_two_stage_sampling}). 
Substituting this definition of the mean into the kernel expression yields the 
following weighted centring formula for the entries of the Gram matrix:
\begin{equation}
    \tilde{\mathbf{K}}_{ij}
    \;=\; \mathbf{K}_{ij} 
    - \sum_{l=1}^{n} w_l\,\mathbf{K}_{il}
    - \sum_{l=1}^{n} w_l\,\mathbf{K}_{lj} 
    + \sum_{l=1}^{n}\sum_{m=1}^{n} w_l w_m\,\mathbf{K}_{lm}.
    \label{eq: weighted_centered_gram}
\end{equation}
The four terms in~\eqref{eq: weighted_centered_gram} have an intuitive 
interpretation: the first is the raw similarity $\mathbf{K_{ij}}$, the next two subtract 
the contributions of the mean from each argument, and the final term adds back 
the mean--mean similarity. 
Therefore, applying \eqref{eq: weighted_centered_gram} directly to the unweighted, uncentered Gram matrix is sufficient to obtain the correctly centred version.
The eigen-decomposition of $\tilde{\mathbf{K}}$ provides the principal 
components in the feature space. If $\tilde{\mathbf{K}}\boldsymbol{\alpha} = 
\lambda \boldsymbol{\alpha}$, then the associated principal axis in 
$\mathcal{H}$ is $\mathbf{v} = \sum_{i=1}^{n} \alpha_i 
\tilde{\Phi}(\mathbf{x}_i)$, and the projection of a new sample $\mathbf{X}$ 
onto $\mathbf{v}$ is obtained via the centred kernel
\[
    \tilde{k}(\mathbf{x}_i,\mathbf{x}) 
    = k(\mathbf{x}_i,\mathbf{X})
    - \sum_{l=1}^{n} w_l\,k(\mathbf{x}_l,\mathbf{X})
    - \sum_{l=1}^{n} w_l\,k(\mathbf{x}_i,\mathbf{x}_l)
    + \sum_{l=1}^{n}\sum_{m=1}^{n} w_l w_m\,k(\mathbf{x}_l,\mathbf{x}_m).
\]

For the~\gls{ae} model, the adjustment of the loss function follows directly from 
the observation that the reconstruction error is given by the~\gls{mse} between the original input $\mathbf{x}$ and its reconstruction obtained via the 
composition of the encoder $f_{\theta}$ and decoder $g_{\varphi}$. 
Explicitly, the 
standard loss takes the form
\[
    \mathcal{L}_{\text{AE}}(\mathbf{x}; \theta, \varphi) 
    \;=\; \frac{1}{n} \sum_{j=1}^{n} 
    \big\| \mathbf{x}_j - g_{\varphi}(f_{\theta}(\mathbf{x}_j)) \big\|^2.
\]
Since the~\gls{mse} corresponds to an average, it is natural and legitimate to apply the adjusted averaging scheme introduced in ~\eqref{eq: adjusted_average_two_stage_sampling}. Incorporating the sample weights 
$w_j$, the weighted autoencoder loss becomes
\begin{equation}
    \tilde{\mathcal{L}}_{\text{AE}}(\mathbf{x}; \theta, \phi) 
    \; \frac{1}{n} \; \sum_{j=1}^{n} w_{j}\,
    \big\| \mathbf{x}_j - g_{\phi}(f_{\theta}(\mathbf{x}_j)) \big\|^2,
    \label{eq: weighted_autoencoder_loss}
\end{equation}
\\
This adjustment ensures that the optimisation process accounts explicitly for the 
sampling design, thereby aligning the training objective with the target population 
distribution. 
As a result, the learned representations and reconstructions are not 
only optimise for the observed sample, but are also valid and representative for 
population-level inference.

\section{Simulation Study}
\label{apx:simulation_allocation}

This appendix presents the simulation study presented in Section~\ref{sec: Intergative_Sampling}.

\subsection{Population and Domain Construction}

\begin{figure}[H]
    \centering
    \includegraphics[width=0.5\linewidth]{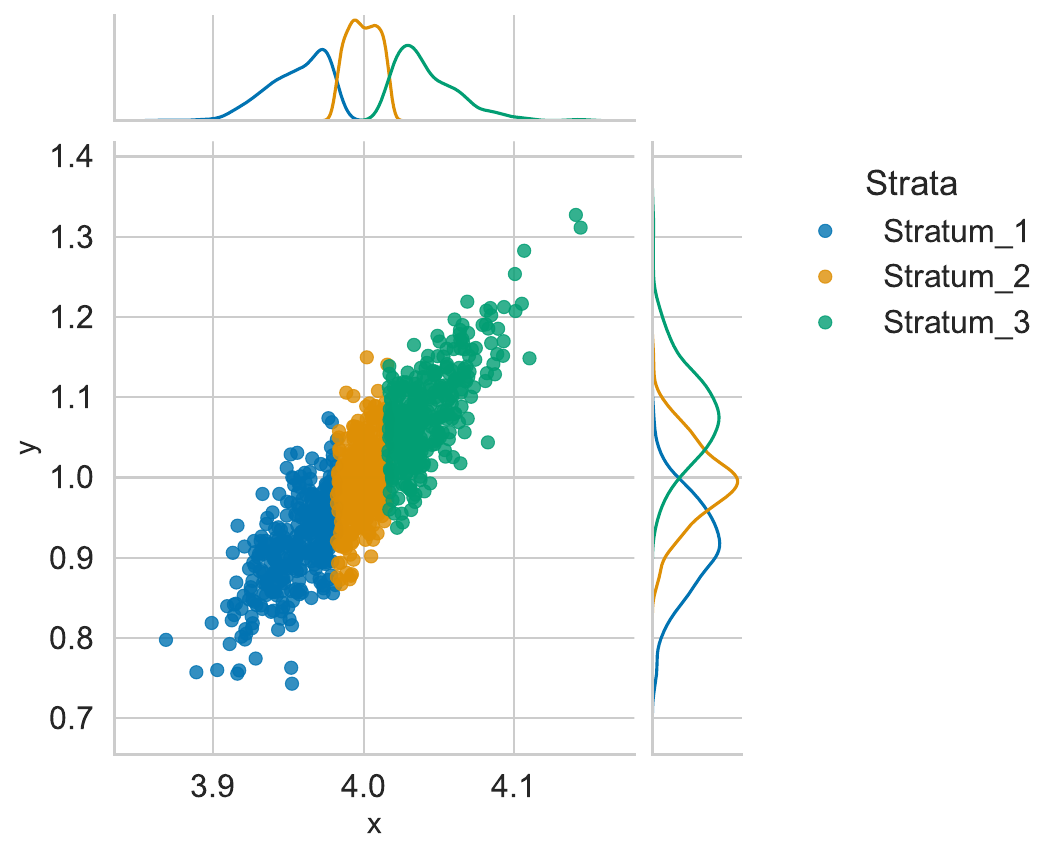}
    \caption{Scatter plot of the synthetic bivariate Gaussian population, stratified by sampling frame. Marginal distributions of each variable are shown along the axes for each stratum.}
    \label{fig: synthetic_population}
\end{figure}

A synthetic population of $N = 3000$ units was generated from a bivariate Gaussian 
distribution with mean vector $\mu = [\mu_y, \mu_x]$ and covariance matrix
\[
\Sigma = \begin{bmatrix} 
\sigma_y^2 & \rho\sigma_y\sigma_x \\ 
\rho\sigma_y\sigma_x & \sigma_x^2 
\end{bmatrix},
\]
truncated to positive values.
Parameter values were set to $\mu_x = 4$, $\mu_y = 1$, 
$\sigma_x = 0.3$, $\sigma_y = 0.2$, and $\rho = 0.85$. 
A visual depiction of the synthetic population is provided in 
Figure~\ref{fig: synthetic_population}.

The $y$-variable was treated as the survey response. 
For the multi-frame setting, units were 
sequentially allocated into $D = 2^Q - 1$ disjoint domains, according to 
pre-specified domain sizes. Domains were subsequently aggregated into $Q = 3$ 
frames based on inclusion indicators, ensuring that the union of frames reproduced the full population while allowing for controlled frame overlap. 
In contrast, for the~\gls{sts} setting, we also considered $Q = 3$ strata, in order to facilitate interpretability and comparability of results across the two designs. 

\subsection{Sampling and Allocation}

The two sampling strategies differed in both design and implementation. 
For the~\gls{sts} case, the most straightforward setting, simple random sampling without replacement, was employed within each stratum. 
By contrast, in the multi-frame setting, Sampford sampling was used to select units within each frame, 
ensuring valid inclusion probabilities irrespective of estimator choice. 
Two allocation schemes were examined. In the proportional allocation scheme, 
sample sizes were assigned in proportion to frame sizes, 
\[
n_q = n \cdot \frac{N_q}{\sum_{q=1}^Q N_q}, \qquad \text{ for } q=1,\ldots, Q
\]
where $n$ is the overall sample size and $N_q$ the population size of frame $q$. 
This allocation is computationally simple and effective when the variability across frames is homogeneous.  

The second scheme was a cost-optimal allocation under a fixed total survey budget $C$, with a linear cost function
\[
C = c_0 + \sum_{q=1}^Q n_q c_q,
\]
where $c_0$ denotes fixed costs and $c_q$ the per-unit cost of surveying frame $q$. 
The optimal frame-specific sample sizes are
\[
n_q^{opt} = \frac{(C-c_0) N_q \sigma_{q\alpha} / \sqrt{c_q}}{\sum_{q=1}^Q N_q \sigma_{q\alpha} \sqrt{c_q}},
\]
where $\sigma_{q\alpha}$ is the frame-specific standard deviation adjusted for 
unit multiplicity. 
This allocation maximises estimator precision given the cost 
constraint \citep{Lohr&Brick14, Sarndal&al92}. In practice, $\sigma_{q\alpha}$ may 
be estimated from historical data, expert judgment, or pilot studies. 
For convenience, we assumed no fixed cost ($c_0 = 0$) and considered frame-specific 
per-unit costs that were approximately equal but with slight differences. 
Specifically, we set $c_1 = 9.5$, $c_2 = 10$, and $c_3 = 10.5$. 
For the stratified case, the multiplicity adjustment is not needed, leading to the simplified 
expression \citep{lohr2021sampling},
\[
n_q^{opt} = \frac{(C-c_0) N_q \sigma_{q} / \sqrt{c_q}}{\sum_{q=1}^Q N_q \sigma_{q} \sqrt{c_q}},
\]
where $\sigma_q$ is the within-stratum standard deviation. In realistic applications, 
frame or stratum variances are rarely known a priori and must be estimated. 
In the simulation setting, however, the true values were employed, since the entire synthetic population was available.

\subsection{Estimation Methods}

For the multi-frame design, two estimators were evaluated. 
The first one was the Simple Multiplicity (SM) estimator, which adjusts for unit multiplicity $m_{kq}$, 
\[
\hat  Y_{SM} =  \sum_{q=1}^Q  \sum_{k \in s_q} y_k   m_{kq}^{-1} \pi_{kq}^{-1},
\]
where $\pi_{kq}$ is the first-order inclusion probability, and units in the same domain share the same multiplicity; 
$m_{kq}$ denotes the multiplicity of the $k$-th individual within the $q$-th frame; see \cite[]{Mecatti&Singh14}.  

The second estimator was the Pseudo Maximum Likelihood 
(PML) estimator, which accounts for frame overlap and domain allocation by iteratively 
adjusting domain-frame totals.
This method, originally proposed by \cite{lohr2006multiple}, provides an optimal linear estimator of the population total when samples are drawn independently from each frame using probability sampling designs. For a detailed discussion of the optimisation procedure, the reader is referred to \cite{lohr2006multiple}.

For the stratified design, the classical stratified estimator was applied, which is defined as the weighted sum of within-stratum means, with weights equal 
to the population stratum proportions.

\subsection{Monte Carlo Evaluation}

Each simulation scenario was replicated $N_\text{run}$ times, using a fixed random 
seed to ensure reproducibility. Performance was assessed through the~\gls{mse} 
(see Section~\ref{sec:simulativescenarios}). 
For each estimator and allocation scheme, mean estimates and 95\% bootstrap 
confidence intervals were reported. Additional evaluation criteria, such as 
variance reduction and cost efficiency, are discussed in the main text. 

\end{appendices}

\section*{Code Avaliability}
Python code available at \url{https://github.com/glancia93/Data_Driven_Strategies_for_Detecting_and_Sampling_misrepresented_subgroups}

\end{document}